\documentclass[journal, 12pt,draftcls,onecolumn,twoside]{IEEEtran}

\normalsize

\usepackage{cite}
\ifCLASSINFOpdf
 
\else
   \usepackage[dvips]{graphicx}
 \fi
\usepackage[cmex10]{amsmath}
\usepackage{amssymb}
\usepackage{stfloats}
\usepackage{amsthm} 
\usepackage{wrapfig}
\usepackage{latexsym}
\usepackage{bm, bbm} 
\usepackage{epsfig} 
\usepackage{graphicx}
\usepackage{epsfig}
\usepackage{multicol}
\usepackage{psfrag}
\usepackage{float}
\usepackage{subfigure}
\usepackage{color}

\newcommand{\Z}{\mathbb{Z}}
\newcommand{\Ftwo}{\mathbb{F}_2} 

\newcommand{\matr}[1]{{#1}}

\newcommand{\tr}{\mathsf{T}}

\newcommand{\defeq}{\triangleq}
\newcommand{\cC}{\mathcal{C}}
\def\girth{{\rm girth}}

\newtheorem{lemma}{Lemma}
\newtheorem{theorem}[lemma]{Theorem}

\newtheorem{corollary}[lemma]{Corollary}

\theoremstyle{plain}

\newtheorem{PreDefinition}[lemma]{{\textbf{Definition}}}
    {\begin{PreDefinition}}{\hfill$\square$\end{PreDefinition}}

\theoremstyle{plain}

\newtheorem{PreRemark}[lemma]{{\textbf{Remark}}}
  \newenvironment{remark}%
    {\begin{PreRemark}\upshape}{\hfill$\square$\end{PreRemark}}

\newtheorem{PreExample}[lemma]{{\textbf{Example}}}
  \newenvironment{example}%
    {\begin{PreExample}\upshape}{\hfill$\square$\end{PreExample}}

\usepackage{tikz}
\newcommand*\circled[1]{\tikz[baseline=(char.base)]{
  \node[shape=circle,draw,inner sep=1pt] (char) {#1};}}

\usepackage{xcolor}
\newcommand{\shade}[1]{%
 \colorbox{red!20}{$\displaystyle#1$}}

 \newcommand\red[1]{{\color{red}#1}}
  \newcommand\blue[1]{{\color{blue}#1}}
\newcommand\green[1]{{\color{green}#1}}
\newcommand\orange[1]{{\color{orange}#1}}

\begin{document}

\title{A Unifying Framework to Construct\\  QC-LDPC Tanner Graphs 
of  Desired Girth}
\author{Roxana~Smarandache,~\IEEEmembership{Senior Member,~IEEE,} and \\David~G.~M.~Mitchell,~\IEEEmembership{Senior Member,~IEEE.}
\thanks{This material is based upon work supported by the National Science Foundation under Grant Nos. OIA-1757207 and HRD-1914635. This paper was presented in part at the 2021 IEEE International Symposium on Information Theory \cite{sm21}.}
\thanks{R.~Smarandache is  with the Departments
of Mathematics and Electrical Engineering, University of Notre Dame, Notre Dame, IN 46556, USA. D.~G.~M.~Mitchell is with the Klipsch School of
Electrical and Computer Engineering, New Mexico State University, NM 88003, USA
(e-mail: rsmarand@nd.edu;~dgmm@nmsu.edu).}}
\markboth{IEEE Transactions Information Theory (Submitted paper)}%
{Submitted paper}



\maketitle
\begin{abstract}
 This paper presents a unifying framework to construct low-density parity-check (LDPC) codes with associated Tanner graphs of desired girth. Towards this goal, we  highlight the role that a certain square matrix that appears in the product of the parity-check matrix with its transpose has in the construction of codes with graphs of  desired girth and further explore it in order to generate the set of necessary and sufficient conditions for a Tanner graph  to have a given girth between 6 and 12.   For each such girth, 
 we present algorithms to construct codes of the desired girth and we show how to use them to compute the minimum necessary value of  the lifting factor.  For girth larger than 12, we show how to use multi-step graph lifting methods to deterministically  modify codes in order to increase their girth.   We also 
give a new perspective on LDPC protograph-based parity-check matrices by viewing them as  rows of a parity-check matrix equal to the sum of certain  permutation matrices 
and obtain an important connection between all protographs and those with variable nodes of degree 2. 
We also show that the results and methodology that we develop for the all-one protograph can be used and adapted to analyze the girth of the Tanner graph of any parity-check matrix and demonstrate how this can be done 
using a well-known irregular, multi-edge protograph specified by the NASA Consultative Committee  for  Space  Data  Systems  (CCSDS). 
Throughout the paper, we exemplify our  theoretical results with constructions of LDPC codes with Tanner graphs of any girth between 6 and 14 and   give sufficient conditions for a multi-step lifted parity-check matrix  to have girth between  14 and 22. 
 \end{abstract}

\IEEEpeerreviewmaketitle

\section{Introduction}\label{sec:intro}
Low-density parity-check (LDPC) codes, in particular quasi-cyclic LDPC (QC-LDPC) codes, are now found in many industry standards. One of the main advantages of QC-LDPC codes is that they can be described simply, and as such are attractive for implementation purposes since they can be encoded with low complexity using simple feedback shift-registers \cite{lcz+06} and their structure leads to efficiencies in decoder design \cite{wc07}. The performance of an LDPC  code with parity-check matrix $H$ depends on cycles in the associated Tanner graph, since cycles in the graph cause correlation during iterations of belief propagation decoding \cite{ru01}. Moreover, these cycles form substructures found in the undesirable trapping and absorbing sets that create the  error floor. Cycles have also been shown to decrease the upper bound on the minimum distance (see, e.g., \cite{sv12}). Therefore, codes with large girth are desirable for good performance (large minimum distance and low error floor).  Significant effort has been made to design QC-LDPC code matrices with large minimum distance and girth, see \cite{klf01,tss+04,kncs07,phns13,kb13,msc14,7398027,9241207} and references therein.

In this paper, we will use some previous results by McGowan and Williamson \cite{mw03} and the terminology introduced in  Wu et al. \cite{wyz08}  that  elegantly relate  the girth of $H$ with the girth of $\matr{B}_t(H)\defeq \left(\matr{H}\matr{H}^\tr\right)^{\lfloor{t/2}\rfloor}\matr{H}^{(t\mod 2)}$, $t\geq 1$.  
We take  this connection further and showcase a submatrix $C_H$   of $HH^\tr$  of relevance when looking for  cycles in the Tanner graph of $H$. 
Specifically, we show that the girth of a Tanner graph of an  $n_cN\times n_vN$ parity-check matrix $H$ based on the $(n_c, n_v)$-regular fully connected (all-one) protograph, with lifting factor $N$,  is directly related to the properties of the product $C_H^{\lfloor{t/2}\rfloor}\matr{H}^{(t\mod 2)}$, $t\geq 1$,  and then 
 further exploit this connection 
in order to give the sets of necessary and sufficient conditions for a QC-LDPC code based on an $n_c\times n_v$ all-one protograph, with $n_c=2,3,$ and $4$, respectively,  to have any girth between 6 and 12. These conditions are then used 
to write fast algorithms to construct codes of such desired girth. 
We also  show that, when constructing parity-check matrices of girth larger than $2l$, we need to consider all $l\times n_v$ submatrices  and impose the girth conditions on them.  In particular,  if we want to construct $n_cN\times n_vN$ protograph-based parity-check matrices of girth larger than  $4$, $6$, and $8$, respectively, it is necessary and sufficient to give the  girth conditions for the cases $n_c=2$, $3$, and $4$, respectively.  It follows from these observations that the cases $n_c=2$, $3$, and $4$ that we consider in this paper are not just particular cases, but provide the girth framework for the $n_c\times n_v$ all-one protograph, for {\em all} $n_c\geq 2$. 

We also 
give a new perspective on $n_cN\times n_vN$ LDPC protograph-based parity-check matrices by viewing them as $n_cN$ rows of a parity-check matrix equal to the sum of certain $n_vN\times n_vN$ permutation matrices. Together with our results  that 
the cycles in the Tanner graph  of a  $2N\times n_vN$ parity-check matrix $H$ based on the $(2, n_v)$-regular fully connected (all-one) protograph correspond one-to-one to the cycles in the Tanner graph of a $N\times N$ matrix $C_{12}$,  obtained from $H$ by adding certain permutation matrices in its composition, we obtain an important connection between $n_c\times n_v$ protographs, for any $n_c\geq 2$,  and protographs with check-node degree $n_c=2$. Therefore, although the case of $2 \times  n_v$ protographs  seems of limited practical importance on its own or important only  as part of a larger protograph, with this above-mentioned  new perspective,  it is in fact relevant in connection to any $n_c\times n_v$ protograph.  In addition, square $n_vN\times n_vN$ parity-check matrices equal to sums of  permutation matrices have enjoyed a lot of attention in the context of projective geometry codes~\cite{klf01,lkf02,sv07},  so they are important as well.

Although we mostly assume the case  of  an  $(n_c, n_v)$-regular fully connected protograph,   the results and methodology can be used and adapted to analyze the girth of the Tanner graph of any parity-check matrix.  We exemplify how this can be done 
using an irregular, multi-edge protograph (with entries $0$, $1$, $2$ rather than just $1$ as in the all-one protograph)  specified by the NASA Consultative Committee  for  Space  Data  Systems  (CCSDS) \cite{ccsds12,ddja09}. Here, we show how to obtain the matrix $C_H$ 
and how the results from the  all-one protograph can be 
adapted  to this type of protograph to give 
the necessary and sufficient girth conditions. 

 We also extend our results and methodology  to obtain codes with girth larger than 12. QC-LDPC Tanner graphs directly circularly lifted from a protograph containing a $2\times 3$ all-one sub-protograph, 
referred to as a QC lifting, cannot be considered anymore, see, e.g., \cite{sv12},  
therefore, we need to consider a matrix composed of permutation matrices  such that some are not circulant. 
In order to obtain an increase in girth beyond the restrictive upper bound 12 (and/or increased minimum distance),  
we demonstrate how a deterministic multi-step graph lifting approach, called pre-lifting \cite{msc14}, can be applied. 
Towards this goal, we first  show that any $n_cN\times n_vN$ circularly lifted graph (defining an arbitrary QC-LDPC code) with $N=N_1N_2$ is equivalent to a graph derived from a $n_c\times n_v$ protograph, circularly pre-lifted with a (first) lifting factor  $N_1$ and then circularly lifted  with a (second) lifting factor equal to  $N_2$. We then show and exemplify for $n_c=3$ that 
graphs of $n_cN\times n_vN$ parity-check matrices can be pre-lifted in a deterministic way  in order to increase their girth and/or minimum distance, whereby exponents are modified to break the (circular) limiting  structure of the original QC-LDPC code. We used this approach to construct QC-LDPC codes of girth 14 starting from certain QC-LDPC codes of girth 10 or 12 that we deterministically choose such that their equivalent structures in which the pre-lifts are observed  allow for  slight modifications of  the exponents to yield a girth increase. 
We also give sufficient conditions for a pre-lift of to allow for girth from between 14 and 22. 

 The structure of the paper is as follows. Section~\ref{sec:background} 
 contains  the background results and needed terminology, while  
 Section~\ref{sec:mbyn} 
exploits  these results to show how the girth of  $H$  is directly related to the properties of the product $C_H^{\lfloor{t/2}\rfloor}\matr{H}^{(t\mod 2)}, t\geq 1$,  
and how this connection  can be used  to obtain necessary and sufficient conditions on  $H$ to have  girth larger  than $4$, $6$, and 8.   
In Sections~\ref{sec:2byn} and~\ref{sec:3bynv}
we give the sets of necessary and sufficient conditions for a QC-LDPC code based on an $n_c\times n_v$ all-one protograph, $n_c=2,3,4$,  to have any girth between 6 and 12, and use them  
to write fast algorithms to construct codes of such desired girth. 
We also present a new perspective on $n_cN\times n_vN$ LDPC protograph-based parity-check matrices by viewing them as $n_cN$ rows of a parity-check matrix equal to the sum of certain $n_vN\times n_vN$ permutation matrices. 
In Sec.~\ref{sec:girth>12}, exemplified in the case  $n_c=3$, 
we extend our results and methodology  to obtain codes with girth larger than 12 by considering a 2-step lifting method,  and give 
sufficient conditions for a pre-lift in order to allow for girth from 14 to 22. 
In Section~\ref{sec:multi},  we show how to obtain the matrix $C_H$ for an irregular, multi-edge protograph used in the NASA CCSDS  LDPC  code and how to adapt the results for the all-one (regular, single-edge) protograph 
  to this irregular protograph. We then exemplify the modified approach by giving  
the necessary and sufficient conditions  for  this protograph to have girth larger than 4. 
 Section~\ref{sec:simulations} contains computer simulations  of some of these codes, confirming the expected robust error control performance,  while Section~\ref{sec: conclusion} contains concluding remarks. Lastly,  
Appendices~\ref{pre-lift-observed-1} and~\ref{pre-lift-observed-2}  revisit  two of the examples in the paper, Examples~\ref{2by3}  and  \ref{girth12-from 10}, respectively,  
in order to show how the pre-lifting techniques presented in Section~\ref{sec:girth>12} can be used to obtain a girth increase and, possibly, a minimum distance increase.  
 
We emphasize that what we present is a unifying framework, in the sense that every previous construction of codes of a certain girth must fit in this framework, since we provide {\em the} set of  {\em necessary}  and sufficient conditions for a given girth to be achieved. The construction papers so far have given sufficient conditions for  a code to have  Tanner graphs  of a certain girth.   For example, the literature on eliminating 4-cycles in LDPC codes by choosing the exponents from difference sets is large \cite{fos04,1302295,Fujisawa2005ACO,6310171,DAQIN2018786}.  
It is  what we do, and  it is, in fact,  what the Fossorier conditions,  displayed here in  Corollary~\ref{thm:Fossorier},  do as well. The novelty of this paper is that it shows that these are not only sufficient but also necessary conditions in order to get girth 6 and it provides in a simple minimal format all the conditions that the  differences of the exponents must satisfy in order to result in a code of Tanner graph of girth $6$, $8$, $10$, and $12$, respectively.   In addition, the set of minimal conditions for a desired girth allows for our proposed algorithms to choose lifting exponents to be  extremely fast, in fact they can be evaluated by hand.  Lastly, if desired,  they can display codes of a given girth for the smallest graph lifting factor $N$. Therefore,   we do not exhaustively visit other constructions found in the literature  because of this very  different scope of our paper.  Although we mostly assume the case of an $(n_c,n_v)$-regular fully connected protograph, for $n_c = 2, 3$, and $4$, the results can be used to analyze the girth of the Tanner graph of any parity-check matrix. We note that the theory would need to be suitably adapted;  this is addressed and exemplified in Section~\ref{sec:multi}.%

\section{Definitions, notation, and background}\label{sec:background}
As usual, an LDPC code $\cC$  
is described as the null space of a 
parity-check matrix $\matr{H}$ 
 to which we associate a Tanner
graph~\cite{tan81} in the usual way. The girth of the graph of $H$, denoted by $\girth(\matr{H})$, is  the length of the shortest cycle in the graph.  If a matrix has an entry larger than 1 then we say the corresponding graph has multiple edges between a pair of nodes. We say that a graph has girth 2 if it has multiple  edges.

A protograph \cite{tho03,ddja09} is a small bipartite graph  represented by an $n_c\times n_v$ 
biadjacency matrix  $B_{n_c\times n_v}$ with non-negative integer entries $b_{ij}$, which we also refer to as a protograph. The parity-check matrix $H_{n_c}$ (or $H$ when $n_c$ is clear from the context) of an LDPC block code based on the protograph $B_{n_c\times n_v}$ can be created by replacing each non-zero entry $b_{ij}$  by a sum of $ b_{ij}$ non-overlapping $N\times N$ permutation matrices and a zero entry by the $N\times N$ all-zero matrix. Graphically, this operation is equivalent to taking an $N$-fold graph cover, or ``lifting'', of the protograph. We call the resulting code  a {\it protograph-based} LDPC code. 

Throughout the paper, we use, for any positive integer $L$,
the notation $[L]$ to denote the set $\{  1,2, \ldots, L \}$,  while, for any set, we say that it has  {\em maximal size} if all the possible values that can be generated for the set should be distinct. 

A special  notation  used throughout the paper is  the elegant triangle operator 
 introduced in \cite{wyz08} between any two non-negative integers $e,f \in \Z$ to define 
\begin{align*} d&\defeq e\triangle f\defeq
         \begin{cases} 
           1, 
             & \text{if $e\geq 2,  f=0 $} \\
           0,                                     
             & \text{otherwise}
         \end{cases},  \;  
  \end{align*} 
 and between  two $s\times t$ matrices $\matr{E}=(e_{ij})_{s\times t}$ and $\matr{F}=(f_{ij})_{s\times t}$ with non-negative integer entries,  to  define the matrix $\matr{D} =(d_{ij})_{s\times t}\defeq \matr{E}\triangle \matr{F}$ entry-wise as  $d_{ij}\defeq e_{ij}\triangle f_{ij}, \text{ for all } i\in [s], j\in [t].$
%
 
 We denote the $N\times N$ circulant permutation matrix where the entries of the $N\times N$ identity matrix $I$ are shifted to the left by $r$ positions modulo $N$, as $x^r$. Note that $0$ and $1=x^0$ correspond to the all-zero and identity matrices, respectively, where the dimensions are implied by the context. 
 We say that a (permutation) matrix $P$ has {\em a fixed column (or row)}, and  write 
   $\left(\matr{P}+\matr{I}\right) \triangle \matr{0}\neq\matr{0},$\footnote{The matrix addition is performed over $\Z$.} 
  if it overlaps with the identity matrix in at least one column (or row).  It follows that any  two permutation matrices $\matr{P}$ and $\matr{Q}$ have no common column  if and only if  $\left(\matr{P}+\matr{Q}\right) \triangle \matr{0}=\matr{0}\Leftrightarrow \left(\matr{P}\matr{Q}^\tr +I \right) \triangle \matr{0}=\matr{0} \Leftrightarrow\left(\matr{P}^\tr \matr{Q} +I \right) \triangle \matr{0}=\matr{0}\Leftrightarrow \left(\matr{Q}^\tr \matr{P} +I \right) \triangle \matr{0}=\matr{0}\Leftrightarrow \left(\matr{P}^\tr +\matr{Q}^\tr   \right) \triangle \matr{0}=\matr{0}$  where the matrix addition is performed over $\Z$. 
In addition,   $(P+I) \triangle \matr{0} =0 \Leftrightarrow (P^i+I)\triangle \matr{0} =0,$ for all integers $i\geq 1$. Lastly, we state the following property in a lemma, since it will be used repeatedly in our results.  
\begin{lemma}\label{AB} Let  $\matr{A}=(a_{ij})_{s\times t}$ and $\matr{B}=(b_{ij})_{s\times t}$ be two  matrices with non-negative integer entries, then the  equality $(\matr{A}+\matr{B}) \triangle \matr{A}=\matr{B} \triangle \matr{A}$ holds. 
\end{lemma}
\begin{IEEEproof} The claim follows from the entry-wise equalities. 
$$(a_{ij}+b_{ij})\triangle a_{ij} =  \begin{cases} 
           1 
             & \text{if $a_{ij}+b_{ij} \geq 2,  a_{ij}=0$ } \\
           0                                     
             & \text{otherwise}
         \end{cases} \; =  \begin{cases} 
           1 
             & \text{if $b_{ij} \geq 2,  a_{ij}=0$ } \\
           0                                     
             & \text{otherwise}
         \end{cases} \;= b_{ij}\triangle a_{ij}. $$ 
\end{IEEEproof}

The $\triangle $ operator is used also in the following theorem of  \cite{mw03} and  \cite{wyz08}  to describe an important connection between $\girth(H)$ and matrices $\matr{B}_t(H)\defeq \left(\matr{H}\matr{H}^\tr\right)^{\lfloor{t/2}\rfloor}\matr{H}^{(t\mod 2)}, t\geq 1. $ 
This connection forms  the base  of our paper.  

\begin{theorem}(\hspace{-0.01mm}\cite{mw03} and \cite{wyz08})\label{adjacent-cond} A Tanner graph of an LDPC code with parity-check matrix $\matr{H}$ has  $\girth(H)>2l$ if and only if 
 $\matr{B}_t(H)\triangle \matr{B}_{t-2}(H) =\matr{0}, t=2,3,\ldots, l.$   \end{theorem}

 Lastly, we extend the  theorem  on cycles in all-one protographs from \cite{msc14} that gives the algebraic conditions imposed by a  cycle of length $2l$ in the Tanner graph of an all-one protograph-based LDPC code to the more general case of any protograph.  
  
    \begin{theorem}\label{th:cycle}
Let $\cC$ be a code 
   described by a protograph-based  parity-check matrix $\matr{H}$ 
   where each $(i,j)$ entry is the $N\times N$ zero matrix or a sum of $N\times N$ non-overlapping permutation matrices.  
 Then, a cycle of length $2l$ in  the Tanner graph associated with  $\matr{H} $ is a lifted cycle of  a $2l$-cycle in the protograph, i.e., one that visits  sequentially the groups of $N$ copies of check and variable nodes in the same order of the cycle in the protograph. 
 Therefore, the $2l$-cycle is associated with a sequence of  permutation matrices  $ \matr{P}_{i_0j_0}, \matr{P}_{i_1j_0},  \matr{P}_{i_1j_1},  \matr{P}_{i_2j_1}, \ldots,   \matr{P}_{i_{l-1}j_{l-1}},  \matr{P}_{i_0j_{l-1}}$ (with no two equal adjacent permutations)   
 such that 
$\left(\matr{P}_{i_0j_0}\matr{P}_{i_1j_0}^\tr  \matr{P}_{i_1j_1} \matr{P}_{i_2j_1}^\tr\cdots  \matr{P}_{i_{l-1}j_{l-1}} \matr{P}_{i_0j_{l-1}}^\tr +I\right)\triangle 0\neq0.$ 
\end{theorem}
%
 \begin{corollary} \label{thm:Fossorier} Let $\cC$ be a code 
   described by a  parity-check matrix $\matr{H}$ $=(\matr{P}_{i,j})$ $\in \Ftwo^{n_cN \times n_vN}$,
   where each $\matr{P}_{i,j}$ is an $N\times N$  circulant matrix  $x^{s_{ij}}$.  
   Then the Tanner graph associated with  $\matr{H} $  has a cycle of length $2l$
if there exist indices  $i_0,i_1,\ldots,i_{l-1}$ and $j_0,j_1,\ldots j_{l-1}$     such that $i_s\neq i_{s+1} , j_s\neq j_{s+1}$ (where $s+1$ here means $s+1\mod l$), for all $s\in \{0, 1, \ldots l-1\}$, and such that 
$s_{i_0j_0}-s_{i_1j_0}+  s_{i_1j_1} - s_{i_2j_1} +\cdots + s_{i_{l-1}j_{l-1}} -s_{i_0j_{l-1}}=0. $
\end{corollary}


\section{The matrix $C_H$ and the relation between $\girth(C_H)$  and $\girth(H)$} \label{sec:mbyn} 
In this section,  we use  Theorem~\ref{adjacent-cond} in order to highlight a relation that exists between $\girth(H)$ and the girth of a certain submatrix $C_{H}$ of $HH^\tr$.   
From this, 
we obtain the necessary and sufficient conditions for the graph of $H$ to have girth $6,8,10,$ or $12$. 
For simplicity,  we assume the case of an $n_c\times n_v$ all-one protograph. However, the techniques developed here can be applied to any protograph as we exemplify in Section~\ref{sec:multi}.

  Let $H_{n_c}$ and $C_{H_{n_c}}$  be defined as\footnote{We will use  the notation $H$ and $C_H$ when $n_c$ is clear from the context.}
\begin{equation}\label{matrix-n_cbyn_v}
 \matr{H}=\matr{H_{n_c}}=\begin{bmatrix} \matr{P_{11}} &\matr{P_{12}}
 &\cdots &\matr{P_{1n_v}}\\
 \matr{P_{21}} &\matr{P_{22}} &\cdots &\matr{P_{2n_v}}\\ 
 \vdots&\vdots
  &&\vdots\\ 
 \matr{P_{n_c1}} &\matr{P_{n_c2}}
 &\cdots &\matr{P_{n_cn_v}}\end{bmatrix},  C_H\defeq C_{H_{n_c}}\defeq  \begin{bmatrix} 0&C_{12} &\cdots &C_{1n_c}\\ C_{21}&0&\cdots& C_{2n_c}\\\vdots&\vdots&&\vdots\\ C_{n_c1}&C_{n_c2}&\cdots &0\end{bmatrix}, \end{equation}
 where 
\begin{equation}\label{Cij} C_{ij}\defeq C_{ji}^\tr\defeq \matr{P_{i1}} \matr{P_{j1}}^\tr +\cdots+  \matr{P_{in_v}} \matr{P_{jn_v}}^\tr, \quad i,j\in[n_c],   \end{equation}
where $P_{ij}$ are permutation matrices, for all $i\in [n_c], j\in  [n_v]$. In this paper, we later focus on the case that the permutation matrices are circulants or arrays of circulants (Sections~\ref{sec:2byn} and \ref{sec:3bynv}), since this will result in QC-LDPC codes that are attractive in practice; however, the results in this section hold for arbitrary permutation matrices. 

Below, we highlight how the matrix $C_H$ appears in the products $B_t$:
 \begin{align*}&\matr{B}_0(H)=\matr{I}, \quad \matr{B}_1(H)=\matr{H}, \quad \matr{B}_2(H)=\matr{H}\matr{H}^\tr=n_vI+C_H, \quad 
 \matr{B}_3(H)=n_vH+C_HH,   \quad 
 \\ &\matr{B}_4(H)=(n_vI+C_H)^2,  \quad  \matr{B}_5(H)=(n_vI+C_H)^2H,  \quad 
 \matr{B}_6(H)=(n_vI+C_H)^3, \text{ etc.} . 
 \end{align*} 
More generally, the following is true for all $m\geq 1$,  
  \begin{align} \label{B-even-CH}&\matr{B}_{2m}(H)=(C_H)^m +n_v\cdot f(I, C_H, C_H^2, \ldots, C_H^{m-1}), \\\label{B-odd-CH}&\matr{B}_{2m+1}(H)=(C_H)^mH +n_v\cdot f(I, C_H, C_H^2, \ldots, C_H^{m-1})H, 
  \end{align}
where $ f(I, C_H, C_H^2, \ldots, C_H^{m-1})$ is a linear function in $I, C_H, C_H^2, \ldots, C_H^{m-1}$. This and Lemma~\ref{AB} give  the following useful equivalences: 
 \begin{align*} &\matr{B}_{2m}(H)\triangle \matr{B}_{2m-2}(H)=0 \Leftrightarrow (C_H)^m \triangle \left(I+ C_H+ C_H^2+ \cdots +C_H^{m-1}\right)=0,\\
 &\matr{B}_{2m+1}(H)\triangle \matr{B}_{2m-1}(H)=0 \Leftrightarrow (C_H)^mH \triangle \left(H+ C_HH+ C_H^2H+\cdots +C_H^{m-1}H\right)=0.\end{align*} 
Therefore, Theorem~\ref{adjacent-cond} can now be restated.
\begin{theorem} \label{adjacent-cond-new} A Tanner graph of an LDPC code with parity-check matrix $\matr{H}$ has  $\girth(H)>2l$ if and only if, for all 
 $t=2,3,\ldots, l$, 
$$(C_H)^{\lfloor \frac{t}{2}\rfloor} H^{(t\mod 2)} \triangle \left(I+ C_H+ C_H^2+ \cdots +C_H^{{\lfloor \frac{t}{2}\rfloor}-1}\right)H^{(t\mod 2)} =0.$$
 \end{theorem}  
 
 In particular, we have the following theorems. 
  \begin{theorem}\label{mbyn-girth>4}  Let $H$, $C_H$, and  $C_{ij}$ be defined as in \eqref{matrix-n_cbyn_v} and \eqref{Cij}.  Then $ \girth(H)>4$ if and only if     
$C_{ij}\triangle 0 =0,$ for all  $ i,j\in[n_c], i\neq j$. Equivalently, $ \girth(H)>4$ if and only if  $C_{ij}$ does not have multiple edges, for all  $i, j\in[n_c], i\neq j$, and equivalently,  if and only if $\girth(C_H) >2$. 
 \end{theorem} 
 \begin{IEEEproof}
 From Theorem~\ref{adjacent-cond-new},  we have the following equivalence   $\girth(H)>4\Longleftrightarrow 
 C_H\triangle I=0\Longleftrightarrow C_{ij}\triangle 0 =0\Longleftrightarrow$  $C_{ij}$ does not have multiple edges, for all  $i, j\in[n_c], i\neq j$.
 \end{IEEEproof} 
  \begin{theorem}\label{mbyn-girth>6}  Let $H$, $C_H$, and  $C_{ij}$ be defined as in \eqref{matrix-n_cbyn_v} and \eqref{Cij}. Then $\girth(H)>6$ if and only if $C_H\triangle 0 =0 $ and 
$  C_HH\triangle H=0$. Equivalently,  $ \girth(H)>6$ if and only if  $ C_{ij}\triangle 0 =0 $ and $
\sum\limits_{{l=1}\atop {l\neq i}}^{n_c} C_{il}P_{lk}  \triangle P_{ik} =0,$ for all $k \in[n_v], i,j\in[n_c], i\neq j.$  \end{theorem}
\begin{IEEEproof} The condition 
 $C_HH\triangle H=0$ is equivalent to  
$\sum\limits_{{l=1}\atop {l\neq i}}^{n_c} C_{il}P_{lk}  \triangle P_{ik} =0,$ for all $k \in[n_v], i\in[n_c]. $ The claim follows. 
\end{IEEEproof} 

\begin{theorem}\label{mbyn-girth>8}  Let $H$, $C_H$, and  $C_{ij}$ be defined as in \eqref{matrix-n_cbyn_v} and \eqref{Cij}.  Then $ \girth(H)>8$ if and only if 
  $ C_H\triangle 0=0$  and  $C_H^2\triangle (I+C_H)=0. $ Equivalently,  $ \girth(H)>8$ if and only if $ C_{ij}\triangle 0 =0 $ and
$\sum\limits_{{l=1}\atop {l\neq j}}^{n_c} C_{il}C_{lj} \triangle C_{ij}=0,$ for all $i,j\in [n_c], i\neq j.$ Moreover, if $n_c=3$,   then $ \girth(H)>8$ if and only if $ \girth(C_H) >4$.\footnote{Due to Lemma~\ref{C-bounded-girth}, we obtain that, for $n_c=3$,  $ \girth(H)>8$ if and only if $ \girth(C_H)=6$.}
\end{theorem}
  \begin{IEEEproof} 
We need to show that by satisfying  $C_H^2\triangle (I+C_H)=0 $ 
we also obtain $C_HH\triangle H=0$. 
Indeed, \\
$\begin{bmatrix} C_{i1} &\hspace{-0.5mm}\cdots\hspace{-0.5mm}& C_{i,i-1} &0& C_{i,i+1} &\cdots & C_{in_c}\end{bmatrix} \begin{bmatrix} C_{i1} & \cdots& C_{i,i-1} &0& C_{i,i+1} &\cdots & C_{in_c}\end{bmatrix}^\tr \triangle I=0 
\Longleftrightarrow 
\sum\limits_{{l=1}\atop {l\neq i}}^{n_c}   C_{il}C_{li} \triangle I=0 \Longleftrightarrow  \sum\limits_{{l=1}\atop {l\neq i}}^{n_c}   C_{il}\left( \sum\limits_{k=1}^{n_v} P_{lk} P_{ik}^\tr\right) \triangle I=0
\Longrightarrow 
\sum\limits_{{l=1}\atop {l\neq i}}^{n_c}  C_{il} P_{lk}P_{ik}^\tr  \triangle I=0 \Longleftrightarrow $\\ $ \sum\limits_{{l=1}\atop {l\neq i}}^{n_c}  C_{il} P_{lk}  \triangle P_{ik}=0, \text{ for all } k\in [n_v] , i\in [n_c],$ which  is equivalent to $B_3(H)\triangle B_1(H) =0$.  

For $n_c=3$, 
    $C_H^2\triangle (I+C_H)=0$ implies that, for all $ i\in [3]$, $\sum\limits_{{l=1}\atop {l\neq i}}^{3} C_{il}C_{li} \triangle I=0$, and,  equivalently, based on Theorem~\ref{adjacent-cond},  that $\girth\begin{bmatrix} C_{12} &C_{13} \end{bmatrix} >4,$ $\girth\begin{bmatrix} C_{21} &C_{23} \end{bmatrix} >4,$ and $\girth\begin{bmatrix} C_{31} &C_{32} \end{bmatrix} >4$.  However, since for a matrix with $n_c=3$,  a 4-cycle occurs in $C_H$ if and only if it occurs in a row of $C_H$, we obtain that $\girth(C_H)>4$. Reversely, if  $\girth(C_H)>4$, then  $C_H^2\triangle I=0$ which also implies the weaker condition    $C_H^2\triangle (I+C_H)=0$. Therefore, for $n_c=3$, we obtain that  $C_H^2\triangle (I+C_H)=0$ is in fact equivalent to $C_H^2\triangle I=0$. \end{IEEEproof} 
Similar theorems can be stated for girth larger than 10, 12, and so on.
 
We exemplify the notions and the results  for $n_c=3$, by revisiting the  following $3N\times 4N$  protograph-based code of girth 10 that can be found in~\cite{msc12,msc14}.  
 
  \begin{example} \label{firstprelift}
 Let  $\matr{H}=\begin{bmatrix}\matr{P_{11}}& \matr{P_{12}}&\matr{P_{13}}&\matr{P_{14}}\\ \matr{P_{21}}& \matr{P_{22}}&\matr{P_{23}}&\matr{P_{24}}\\ \matr{P_{31}}& \matr{P_{32}}&\matr{P_{33}}&\matr{P_{34}}
\end{bmatrix}\defeq
\left[\begin{array}{cc|cc|cc|cc} 
\matr{1} & 0& \matr{1} &0& \matr{1}&0& \matr{1}&0 \\ 
0& \matr{1} & 0& \matr{1} &0& \matr{1}&0& \matr{1}\\\hline
 \matr{1}& 0&\matr{x}&0&0&\matr{x^{10}}&0&\matr{x^{13}}\\ 
 0& \matr{1}&  0&x^5& \matr{x^{10}}&0& \matr{x^{13}}&0\\  \hline
 \matr{1}& 0&0&\matr{x^{7}}&\matr{x^{11}}&0&\matr{x^2}&0\\
 0& \matr{1}& x^7&0&0& \matr{x^{11}}&0& \matr{x^{4}}
 \end{array}\right].$
 
\noindent Here, the permutation matrices are $N_1\times N_1 = 2\times 2$ arrays of $N_2\times N_2$ circulant matrices, such that $N=N_1N_2$. This double lifting is referred to as \emph{pre-lifting}, and results in a QC-LDPC matrix if the second lifting is circulant \cite{msc14}. The matrices $C_{ij}$ and $C_H$ associated with $H$ are
 \begin{align*}& C_{21}=C_{12}^\tr= \begin{bmatrix} 1+x&x^{10}+x^{13} \\ x^{10}+x^{13} &1+x^5 \end{bmatrix}, 
C_{31}=C_{13}^\tr=\begin{bmatrix}1+ x^2+x^{11}& x^7\\ x^7&1 +x^{4}+x^{11}\end{bmatrix}, \\
&C_{23}=C_{32}^\tr=\begin{bmatrix} 1& x^{-6}+x^{-1}+x^{9}\\ x^{-1}+x^{-2}+ x^{11}&1\end{bmatrix}, 
C_H= \begin{bmatrix} 0&C_{12}&C_{13}\\ C_{21}&0&C_{23}\\ C_{31}&C_{32} &0\end{bmatrix}=
\end{align*}
  \begin{align*}
&{\small \left[\begin{array}{cc|cc|cc} 
 0&0& 1+x^{-1}&x^{-10}+x^{-13}&1+ x^{-2}+x^{-11}& x^{-7}\\ 
 0&0&x^{-10}+x^{-13} &1+x^{-5} &x^{-7}&1 +x^{-4}+x^{-11}\\\hline
 1+x&x^{10}+x^{13} &0&0& 1& x^{-6}+x^{-1}+x^{9}\\
x^{10}+x^{13} &1+x^5&0&0&x^{-1}+x^{-2}+ x^{11}&1\\\hline
1+ x^2+x^{11}& x^7&1& x+x^{2}+ x^{-11}&0&0\\
x^7&1 +x^{4}+x^{11}&x^{6}+x+x^{-9} &1&0&0\end{array}\right]}.\end{align*}

The $3N\times 3N = 6N_2 \times 6N_2$, relatively dense $(8,8)$-regular matrix $C_H$ has girth 6 for  $N_2=27$, for example. Equivalently, the $(3,4)$-regular $H$ has girth 10  for any such $N_2$. Moreover, codes with parity-check matrices that are submatrices of $H$ based on $2\times 4$ all-one (sub-)protographs have girth 12, for these $N_2$.
\end{example}
%
%

 \begin{remark} The matrix $C_H$ is therefore relevant when discussing  the girth  of $H$. In particular, we have observed the following two equivalences 
 
 \fbox{$\girth(H)>4 \Longleftrightarrow \girth(C_H)>2$} and \fbox{$ \text{for } n_c=3, \girth(H)>8\Longleftrightarrow \girth(C_H)>4.$}
 
\noindent   Unfortunately, we cannot keep increasing the  girth of $C_H$ in hope to obtain higher girth for the associated $H$, because,  
  if $n_c\geq 3$, $C_H$ cannot have girth larger  than 6, and if $n_c\geq 4$,   $C_H$ cannot have girth larger  than 4, no matter what lifting or pre-lifting we choose.  The 6-cycles and 4-cycles, respectively, are easy to observe.  We state this fact below as Lemma \ref{C-bounded-girth}.
  \end{remark} 
  \begin{lemma} \label{C-bounded-girth} Let $H$ and $C_H$ be matrices  defined as in \eqref{matrix-n_cbyn_v}.  If $n_c=3$,  then $\girth(C_H)\leq 6$,  and if $n_c\geq 4$, $\girth(C_H)\leq 4$. 
\end{lemma} 
  
The following theorems connect, for some particular cases, codes with parity-check matrix $H_{n_c}$ based on a protograph $B$  of size $n_c\times n_v$ to codes with parity-check matrix $H_{m}$ based on a sub-protograph  of $B$ with size $m\times n_v$, $m<n_c$. This allows one to expand $mN\times n_vN$ protograph-based matrices to $ n_cN\times n_vN$ matrices of the same girth after making sure that the $mN\times n_vN$ protograph-based matrices based on the sub-protograph  of size $m\times n_v$, $m<n_c$ of $B$ have the desired girth.

  We start with a simple observation that formally states the following connection between $C_{H_{n_c}}$ and $C_{H_{n_c-1}}$, because it will prove useful in  obtaining the girth conditions for $n_cN\times n_vN$ codes by starting  from  $(n_c-1)N\times n_vN$ codes of desired girth and adding an extra  row. 
 \begin{lemma} \label{induction} Let $n_c\geq 3$, $C_{H_{n_c}}$, $C_{H_{n_c-1}}$  be defined as in \eqref{matrix-n_cbyn_v}, and $C_{n_c,j}$  be defined as in \eqref{Cij} with $i=n_c$, for all $j\in [n_c-1]$. The following 
 decomposition of $C_{H_{n_c}}$ into two submatrices $X_{H_{n_c-1}}$ that contains information about the first $n_c-1$ rows $H_{n_c-1}$  of $H_{n_c}$ (through  $C_{H_{n_c-1}}$), and $Y_{H_{n_c}}$ that contains information about the new added row in $H_{n_c}$ (through  $C_{n_c}$),  can be observed, 
 $$C_{H_{n_c}}= \underbrace{\begin{bmatrix} C_{H_{n_c-1}} &0_{(n_c-1)\times 1}\\0_{1\times (n_c-1)}&0_{1\times 1}\end{bmatrix}}_{X_{H_{ n_c-1}}} + \underbrace{\begin{bmatrix} 0_{(n_c-1)\times (n_c-1)}&C_{n_c}\\C_{n_c}^\tr&0_{1\times1}\end{bmatrix}}_{Y_{H_{n_c}}},  \text{where~}  C_{n_c}^\tr\defeq \begin{bmatrix} C_{n_c1}&\cdots &C_{n_c,n_c-1} \end{bmatrix}.$$
 \end{lemma}
This lemma can be used to obtain the conditions for girth larger than $g$ for a matrix $H_{n_c}$ with check-node degree $n_c$ when we start from a matrix $H_{n_c-1}$ with check node degree $n_c-1$ of girth larger than $g$ and add an extra row of permutations. For example,  to obtain the conditions for girth larger than 8 for a matrix $H_4$ with $n_c=4$,  we start from a matrix $H_3$ with $n_c=3$ of girth larger than 8 and add an extra row of permutations. The necessary condition for girth larger than 8 is   $C_{H_4}^2\triangle (C_{H_4}+I) =0$, which can be rewritten as 
$\left(X_{H_3} +Y_{H_4}\right)^2\triangle \left(X_{H_3} +Y_{H_4}+ I\right)=0, $ from which we obtain, based on the many zeros in the description of $X_{H_3}^2$, $Y_{H_4}^2$, $X_{H_3}Y_{H_4}$, and $Y_{H_3}X_{H_4}$,  the following three entry-wise conditions (obtained by writing the  condition $C_{H_4}^2\triangle (C_{H_4}+I) =0$ entry-wise): 
\begin{align*}& (C_{H_3}^2+C_4C_4^\tr )\triangle (C_{H_3}+I) =0, \quad  C_{H_3}C_4 \triangle C_4=0, \quad C_4^\tr C_4 \triangle I=0, 
\end{align*} 
where $C_{4}^\tr\defeq \begin{bmatrix} C_{41}&C_{42}& C_{43} \end{bmatrix}.$ If we start from a $3\times n_v$ matrix of girth larger than 8, then $C_{H_3}^2\triangle (C_{H_3}+I) =0$ is already satisfied; 
therefore, from the conditions given by this equality, we only need to record the conditions involving  $C_4$. 
Consequently, Lemma~\ref{induction} can be efficiently used to obtain conditions on the permutation matrices on the 4th row of $H_{4}$.  
In addition, we also have the following connection. 
 \begin{theorem}\label{min-conditions}    Let $H_{n_c}$ be defined as in \eqref{matrix-n_cbyn_v}.   
%
Then $H_{n_c}$ has $\girth(H_{n_c})>2m$ if and only if all $\min(m,n_c) \times n_v$ submatrices of $H_{n_c}$ have girth greater than $2m$, for all $m\geq 2$.  
 \end{theorem} 
 \begin{IEEEproof} If $m\leq n_c$, an  $2m$-cycle cannot involve more than $m$ rows of $H_{n_c}$; a cycle involving $m+1$ rows of $H_{n_c}$, must be of length strictly  larger  than $2m$.  If $m>n_c$, then the claim holds trivially. 
 \end{IEEEproof}
 In particular, this theorem gives the following corollary.  
 
%
%
  
\begin{corollary} \label{inductive-larger} Let $H_{n_c}$ be defined as in \eqref{matrix-n_cbyn_v}. Then

 \begin{itemize} \item  
If $n_c\geq 2$, then $\girth(H_{n_c})>4 $ if and only if all $2\times n_v$ submatrices of $H_{n_c}$ have girth $>4$, and, equivalently, if and only if matrices $C_{ij}$ have no multiple edges, for all $i,j\in [n_c], i\neq j$;
  \item If $n_c\geq 3$,  then $\girth(H_{n_c})>6 $ if and only if all $3\times n_v$ submatrices have girth $>6$, and, equivalently, if and only if all $3\times 3$ submatrices of $H_{n_c}$ have the permanent over $\Z$ of maximum possible weight;
\item If $n_c\geq 4$,  then $\girth(H_{n_c})>8$  if and only if all $4\times n_v$ submatrices of $H_{n_c}$ have girth $>8$; 
  \item If $n_c\geq 5$,  then $\girth(H_{n_c})>10 $ if and only if all $5\times n_v$ submatrices have girth  $>10$. 
\end{itemize} 
\end{corollary} 
Lastly, a weaker  result can be also be stated. 
 
  \begin{corollary}\label{mbyn-girth2n}   Let $H_{n_c}$, $C_{H_{n_c}}$, and  $C_{ij}$ be defined as in \eqref{matrix-n_cbyn_v} and \eqref{Cij}.  Let $m\geq 2$.   
  Then $$\girth(H_{n_c})>2m  \Longrightarrow  \girth(C_{ij})>m, \text{ for all } 1\leq i<j \leq n_c.$$ 
 \end{corollary}
\begin{IEEEproof} In order for the graph of $H_{n_c}$ to have a certain girth $2m$,  we need all its submatrices to have girth at least $2m$, including all its $2\times n$ submatrices, graphs of which have twice the girths of their associated $C_{ij}$ matrices.  
 \end{IEEEproof}
\begin{remark} \label{how-to} It follows that, when constructing parity-check matrices of large girth $2m$, we need to make sure that the associated matrices $C_{ij}$ have  girth at least $m$. In addition, we need to consider all $m\times n_v$ submatrices  and impose the girth conditions on them. 
Thus, we can start from a $2\times n_v$ matrix of girth $2m$ and add one row at a time imposing the girth conditions such that the newly formed matrix maintains the girth $2m$. 
\end{remark}  
%
%
In the next sections,  we will use the above results to construct $H_{n_c}$ of various girths for the cases of   $n_c=2,3,$ and $4$. 
 \section{The girth of  $2N\times n_vN$ matrices $H_2$}
 \label{sec:2byn}
Although the case of a $2\times n_v$ protograph seems of  limited practical importance  on its own, it is essential when seen as part of a larger protograph, since each $n_c\times n_v$ protograph of girth $2m$, with $n_c\geq 3$,  has $n_c\choose 2$  $2\times n_v$ protographs that need to have girth at least $2m$.  In addition, we show in the next theorem that the cycles in the Tanner graph of $H_2$ are in one-to-one correspondence with the cycles of the Tanner graph of a sum of permutation matrices in the composition of $H_2$. Parity-check matrices equal to a sum of some permutation matrices  have enjoyed a lot of attention in the context of projective geometry codes \cite{klf01,lkf02,sv07}.  
  
  
In particular, any $n_cN\times n_vN$ protograph-based LDPC parity-check matrix can be seen as a submatrix  of a square $n_vN\times n_vN$ matrix that is,  in fact, a sum of  permutation matrices of size $n_vN\times n_vN$. For example, the $n_v\times n_v$ all-one matrix is a sum of $n_v$ distinct permutation matrices $1+x+x^2+\cdots +x^{n_v-1}$, each of size  $n_v\times n_v$. These matrices are then  lifted with lifting factor $N$  to obtain permutation matrices $P_1,P_2, \ldots, P_{n_v}$, each of  size $n_vN\times n_vN$, giving a sum of the same size. So any $n_cN\times n_vN$ LDPC monomial  matrix can be seen as a submatrix of such a square matrix $P_1+P_2+ \cdots + P_{n_v}$. 

 The following example demonstrates this important fact:  how a protograph-based parity-check matrix can be decomposed (not uniquely) as a sum of permutation matrices. The code we use is the $(128,64)$  NASA CCSDS standard code \cite{ccsds12}.
 \begin{example} \label{NASA-example} Let $N=16$ and 
\begin{equation}\label{CCSDS}
H_{(128,64)}\defeq 
 \begin{bmatrix}
 \boxed{1}+\boxed{\red{x^{7}}} &\orange{x^{2}}&\green{x^{14}}&\red{x^{6}}& 0&\blue{1}&\shade{x^{13}}&\circled{$1$}\\ 
\circled{$x^{6}$}&\boxed{1}+\boxed{\red{ x^{15}}}&\orange{1}&\green{ x}& \red{1}&0&\blue{1}&\shade{x^7}\\
\shade{ x^{4}}& \circled{$x$}&\boxed{1}+\boxed{\red{x^{15}}}&\orange{  x^{14}}& \green{x^{11}}& \red{1}&0& \blue{ x^{3}}\\
 \blue{1}& \shade{x}& \circled{$x^{9}$}&\boxed{1}+\boxed{\red{x^{13}}}&\orange{ x^{14}}&\green{x}&  \red{1}&0\end{bmatrix}. \end{equation}

\noindent Then $H_{(128,64)}$ is a sum of 8 permutation matrices $P_i$ of size $8N \times 8N$, $N=16$, from which we only take the first $4N$ rows.  Indeed, let $P_1$, $P_2$, $P_3$, $P_4$, $P_5$, $P_6$, $P_7$, and $P_8$ be of size $8N\times 8N$,  obtained by taking the following  $8\times 8$ permutation matrices and then lifting them with circulants of size $N=16$ (circulants are only given for the first 4 rows since the rest are arbitrary) as follows: the identity matrix $1$ lifted with $1$, $1$, $1$, $1$ (boxed in the matrix above), 
the identity matrix $2$ lifted with $x^{7}$, $x^{15}$, $x^{15}$, $ x^{13}$ (boxed in red), the circulant matrix $x$ lifted with  $x^{6}$, $x$, $x^{9}$, $1$ (circled), the circulant matrix $x^2$ lifted with   $x^{4}$, $x$, $x^{13}$, $x^7$ (shaded red), the circulant matrix $x^3$ lifted with $1$, $1$, $1$, $x^3$ (blue), the circulant matrix $x^5$ lifted with $x^6$, $1$, $1$, $1$ (red), the circulant matrix $x^6$ lifted with $x^{14}$, $x$, $x^{11}$, $x$ (green), and the circulant matrix $x^7$ lifted with $x^{2}$, $1$, $x^{14}$, $x^{14}$ (orange), respectively. 

Then $H_{(128,64)}=\sum\limits_{i=1}^8 P^\prime_i$, where $P^\prime_i$ is the matrix formed by the  first $4N$ rows of $P_i$, and thus  is a submatrix of a sum of $n_v = 8$ permutation matrices.  We will revisit this matrix in Examples~\ref{NASA-example-cont} and \ref{multipleedges-example}. 
 \end{example}  
  
 We conclude from the above that the case of a $2\times n_v$ protograph is important in its own right, and  will explore  some of the above-mentioned facts below.

%
%
\begin{theorem} \label{2byn} Let 
\begin{equation} \label{H,C}\matr{H}_2=\begin{bmatrix} \matr{I} &\matr{I} &\cdots &\matr{I}\\ \matr{P_1}& \matr{P_2}&\cdots &\matr{P_{n_v}} 
\end{bmatrix}, \quad C_{21}=C_{12}^\tr\defeq\sum\limits_{i=1}^{n_v}\matr{P_i}. \end{equation} 
Then,
 $$\girth(\matr{H_2}) =2\ \girth(C_{12}).$$   
\end{theorem}
\begin{IEEEproof} We show that any cycle of size $2l$ in $H_2$ corresponds one-to-one to a cycle of size $l$ in $C_{12}$. Indeed, from Theorem  \ref{th:cycle}, the Tanner graph associated with $\matr{H}_2 $
has a cycle of length $2l$
if and only if there exist indices  $i_1,i_2,\ldots,i_{l}\in [n_v],$     such that $i_s\neq i_{s+1} $   and 
such that  
$ \matr{I}\matr{P}_{i_1}^\tr  \matr{P}_{i_2} \matr{I}^\tr \matr{I}\matr{P}_{i_3}^\tr \matr{P}_{i_4} \matr{I}^\tr \cdots  \matr{P}_{i_{l-1}}^\tr\matr{P}_{i_{l}} \matr{I}^\tr \triangle I \neq 0\Longleftrightarrow
\matr{P}_{i_1}^\tr  \matr{P}_{i_2} \matr{P}_{i_3}^\tr \matr{P}_{i_4} \cdots \matr{P}_{i_{l-1}}^\tr \matr{P}_{i_{l}}  \triangle I \neq 0.$
  Equivalently, there exist $m_1,m_2, \ldots, m_{l}$ such that 
$ P_{i_1}(m_{2}, m_{1})= P_{i_2} (m_{2}, m_{3})= P_{i_3} (m_{4}, m_{3})= \cdots=P_{i_{l}}(m_{l}, m_{1})=1,$ 
which is equivalent to the existence of an $l$-cycle in $C.$  
\end{IEEEproof}

Since a $2l$-cycle in $H_2$   is equivalent to an $l$-cycle in $C_{12}$, and any bipartite graph can only have even size cycles, $l$ must be even, leading to  the  $2l$-cycle in $H_2$ to have the size a multiple of 4.  Therefore, the girth of a $2\times n_v$ parity-check matrix $H_2$  must be  multiple of $4$.

\begin{corollary} \label{cor-2byn} Let 
$$H_2=\begin{bmatrix}\matr{P_1} &P_2&\cdots &\matr{P_{n_v}}\\  \matr{Q_1}&Q_2&\cdots &\matr{Q_{n_v}} 
\end{bmatrix}, \quad C_{21}\defeq C_{12}^\tr\defeq \sum\limits_{i=1}^{n_v} \matr{P_i^\tr Q_i}.$$ 
Then  $$ \girth(H_2) =2\ \girth(C_{21}). $$
\end{corollary} 
\begin{IEEEproof}  The graph of $H_2$ is equivalent to the graph of the matrix 
$ \begin{bmatrix} \matr{I} &\cdots &\matr{I}\\ \matr{P_1^\tr Q_1}& \cdots &\matr{P_{n_v}^\tr Q_{n_v}}\end{bmatrix}$ which, based on Theorem~\ref{2byn}  has twice the girth of $C_{21}$. 
 \end{IEEEproof} 

The following corollary is a well known fact, see for example~\cite{sv12} for more details.  We state it here because it can be seen as another  corollary of the results regarding the $2\times n_v$ protographs. 

\begin{corollary} \label{girth<12} Let $H_{n_c}$ be an LDPC code based on a protograph $B=(B_{ij})_{n_c \times n_v}$, with $B_{ij}\geq0$ integers. If $B$ has a $2\times  3$ submatrix that has all its entries lifted to circulant matrices in $H_{n_c}$, then $\girth(H_{n_c})\leq 12$.
\end{corollary}

\begin{example} \label{2by3} Let $$H_2=\begin{bmatrix} \matr{I} &\matr{I} &\matr{I}\\ \matr{I}& \matr{P_2}&\matr{P_3}\end{bmatrix}.$$ Then the $4l$-cycles in the Tanner graph  of the $2N\times 3N$ matrix   $H_2$ are in one-to-one correspondence with the $2l$-cycles in the Tanner graph of $C_{12}=I+P_2+P_3$. To insure that  $\girth(H_2)=8$ we need to choose matrices $P_2$ and $P_3$ such that  the matrix $I+P_2+P_3$ does not have multiple edges (and thus has girth greater than 2), while in order for $H_2$ to have girth 12, we need to choose $P_2$ and $P_3$ such that the girth of $I+P_2+P_3$  has girth 6.  For example, we can take $P_2$ and $P_3$ to be circulant, equal to $x$ and $x^3$, respectively, and $N=7$. Then the matrix $I+P_2+P_3=1+x+x^3$ corresponds to the $7\times 7 $ parity-check matrix of the cyclic projective code of size 7,  which  has girth 6. Therefore, the corresponding $14\times 21$ matrix $H_2$ has girth 12. 

Any choice  of $P_2$ and $P_3$ where both are circulants restricts  the girth of  $I+P_2+P_3$ to be at most $6$ (see Corollary~\ref{girth<12}); therefore, in order  to obtain a girth of $H_2$  larger than 12,  we need to take $P_2$ and $P_3$ non-circulant. A convenient way to do this is by a 2-step lifting method consisting of a {\em prelifting}, i.e., forming  a square matrix (circular or not) by lifting with $N_1$,  and then lifting it with circulant permutation matrices of size $N_2$, i.e., forming a permutation matrix as a $N_1 \times N_1$ array of $N_2\times N_2$ circulants. For example, the matrices $P_2$ and $P_3$ below are obtained by first lifting each to a $3\times 3$ matrix given by  the circulant permutation matrices $1$ and $ x^2$ with $N_1=3$, respectively, and then applying a second lifting  with circulants $x, x^{13}, x^7$ and $x,x, x^2$, respectively, i.e., 
$$ P_2= \begin{bmatrix}x&0&0 \\0&x^{13}&0\\ 0&0&x^7\end{bmatrix},\> P_3=  \begin{bmatrix}0&x&0 \\0&0&x^2\\ x&0&0\end{bmatrix},\text{ such that } I+P_2+P_3\defeq  \begin{bmatrix} 1+x& x &0\\ 0 &1+x^{13} &x^2 \\ x& 0& 1+x^7\end{bmatrix}.$$ 
The matrix   $ I+P_2+P_3$ has girth  8 for  $N_2=7$, girth 10  if the lifting is  increased to $N_2=11$, and girth 12  if the lifting is increased to $N_2=31$.  Therefore, the $2N\times 3N = 6N_2 \times 9N_2$ parity-check matrix
$\matr{H}_2$ formed with the above  $P_2$ and $P_3$ 
has girth 16, 20, and 24, for $N=7$, $N=11$, and $N=31$, respectively. (As a side note, the matrix $I+P_2+P_3$ has minimum distance 48 for $N=31$.)
 \end{example}
 In Appendix~\ref{pre-lift-observed-1}, we revisit Example~\ref{2by3} to show how the techniques of Sec.~\ref{sec:girth>12} were used in order to  obtain  matrices of girth beyond 12.    

  The following section provides algorithms to construct $2N\times n_vN$ protograph-based codes of various girth $4m$ and, equivalently, to construct sums of $n_vN\times n_vN$ permutation matrices of girth $2m$, $m\geq 1$.  
  
%
%

\subsection{Case of $\girth(H)=4m$, for $m=2,3$} \label{sec:2byn-girth} 
\begin{theorem}\label{2byn-thm:girth}   Let  $H_2$ and $C_{21}$ be defined as in  \eqref{H,C}, and let  $P_j=x^{i_j}$, for all  $j\in [n_v]$, and $i_1=0$. 

\begin{enumerate} 
\item  $\girth(H_2)=4m >4\Leftrightarrow \girth(C_{21})>2$ if and only if  the set
 $\left\{i_j \mid j \in [{n_v}]\right\}$ is of maximal size.\footnote{Recall that by  {\em maximal size}, we indicate that all the possible values that can be generated for the set should be distinct.}

%
%

\item $\girth(H_2)=4m >8$ if and only if  the set
$\left\{ i_j-i_l\mid j,l\in[n_v], j\neq l \right\}$ 
is of maximal size.
\end{enumerate}  \end{theorem} 
The following are two algorithms based on the conditions of Theorem~\ref{2byn-thm:girth} to construct $H_2$ with girth larger than 4 and 8, respectively. They will later be extended to larger protographs and girths.

 \noindent \fbox{\bf Algorithms $A_{2,g>4}$ and $A_{2,g>8}$}\\
  Step 1:  Set  $i_1=0$. Set  $l=1$.\\
    Step 2: Let $l:=l+1$. \\Choose \fbox{$i_{l} \notin \{i_a \mid a\in[l-1]\}$ for $g>4$} or \fbox{$i_{l} \notin \{ i_a+ i_b-i_c\mid  a,b,c\in[l-1]\}$ for $g>8$}. \\
   Step 3:  If $l=n_v$ stop, otherwise, go to Step 2.


%

\begin{example} \label{example-2byn} We construct a $2\times n_v$ protograph-based matrix of girth $4m$, for $m=2, 3$,  following  Algorithms $A_{2,g>4}$ and $A_{2,g>8}$ and choosing the smallest possible exponents at each step, giving
$$H_{2,g>4}= \begin{bmatrix} 1&1&1&1&1&1&1&1\\ 1& x&x^2& x^3 &x^{4}&x^5&x^6&x^7 
 \end{bmatrix}, \ H_{2,g>8}= \begin{bmatrix} 1&1&1&1&1&1&1&1\\ 1& x&x^3&x^7&x^{12}&x^{20}&x^{30}&x^{44}\end{bmatrix}. $$
\noindent The matrix $H_{2,g>4}$, constructed using algorithm $A_{2,g>4}$,  has girth 8 for $N=8$. In this case $C_{21}=1+x+\cdots +x^7$ is the $8\times 8$ all-one matrix of girth 4.  The matrix $H_{2,g>8}$, constructed using algorithm $A_{2,g>8}$, 
  has girth 12  for  $N=77$, for example.  Equivalently, the corresponding $77\times 77$ matrix $C_{21} =  1+ x+x^3+x^7+x^{12}+x^{20}+x^{30}+x^{44}$ has girth 6.
  \end{example}

 The following lemma gives an easy way to choose the next exponent values  such that they are larger than the ones in the forbidden sets.  
 \begin{lemma} \label{2byn-girth12-construction} Let $H_2$ and $C_{21}$ be defined as in \eqref{H,C}.  Let $i_l$  be defined recursively as
$$i_l=1+2i_{l-1}, \>\>i_1= 0, \>\>l\geq 2.$$ 
\noindent Then the Tanner graph of the code with parity-check  matrix $H_2$ has girth 12 for  some $N$ and, equivalently, $C_{21}$ has girth 6.  \end{lemma} 

We exemplify this easy method below. 
\begin{example} The following matrix has girth 12 for $N=73$
$$H_2=\begin{bmatrix} 
1&1&1&1&1&1&1&1\\
1& x           &x^3       &x^7       &x^{15}  &x^{31}   &x^{63}&x^{127}
\end{bmatrix}.$$    
The corresponding $73\times 73$ matrix $C_{21}=1+ x  +x^3       +x^7       +x^{15}  +x^{31}   +x^{63}+x^{127}$
has girth 6.  We reduce the exponents modulo $N=73$ to obtain 
$C_{21}=1+ x  +x^3       +x^7       +x^{15}  +x^{31}   +x^{63}+x^{54}$. Note that this $N$ is not the minimum for which a code can be found, but it can be easily  obtained by hand and it is  faster than Algorithm $A_{2,g>8}$.  

We also note that the component matrix $C_{21}=1+ x  +x^3   +x^7  +x^{15}$ of the denser $C_{21}$ above has girth 6 for $N=25$ while, as expected,  $C_{21}=1+ x  +x^3 $ has girth 6 for $N=7$ (this is the projective code \cite{sv07}). 
\end{example}

 We now review Example~\ref{NASA-example} in view of the connection of Theorem~\ref{2byn}.
\begin{example}\label{NASA-example-cont} Let $P_i$ defined in Example~\ref{NASA-example} and let  $$H_2 \defeq \begin{bmatrix} I&I&I&I&I&I&I&I\\ P_1&P_2&P_3&P_4&P_5&P_6&P_7 &P_8\end{bmatrix}_{256\times 1024}.$$
Theorem~\ref{2byn} says that $\girth(H_2)=2\ \girth(C_{12})$ where $C_{12} =\sum_{i=1}^8 P_i$ and has girth 6. Therefore $H_2$ is matrix of girth 12. By itself, this is not an interesting code; however, the strongly connected $C_{12}$ is, since the parity-check matrix $H_{(128,64)}$ of the NASA CCSDS $(128,64)$ code is contained as a submatrix and thus $\girth(H_{(128,64)})\geq\ \girth(C_{12})$. 
\end{example} 

\begin{remark} While the $C_{12}$ matrices we construct can be invertible, there are cases of $N$ for which the null-space is non-zero, thus giving codes worth considering. For example, the matrix  $C_{21}=1+ x  +x^3 $ has  girth 6 for different values of $N$; however, the values of $N=7,14,$ etc., are needed in order for the code with parity-check matrix $C_{21}$ to have non-zero codewords (due to the fact that  $1+ x  +x^3 $ is a polynomial divisor  of $x^7-1$).  
\end{remark} 
 \subsection{Case of $\girth(H_2)=4m$, for $m\geq 4$} \label{sec:girth12} If we want to obtain matrices $H_2$ of girth larger than 12, then at least one of the matrices   $P_i$ has to be  non-circulant. In our constructions, we follow a prelifting (double lifting) approach, as demonstrated already in Examples \ref{firstprelift} and \ref{2by3}. 
 For example, permutation matrix $P$ below can be seen as the entry 1 in the protograph, first pre-lifted to the $N_1\times N_1 = 3\times 3$ circulant matrix  $x^2$, before the second lifting with circulants $x^a,x^b,x^c$ with lifting factor $N_2 = N/N_1$, i.e.,  
$$P=\begin{bmatrix} 0&x^a&0\\0&0&x^b\\x^c&0&0\end{bmatrix}, \text{ with }x^2=\begin{bmatrix} 0&1&0\\0&0&1\\1&0&0\end{bmatrix}.$$

The following theorem gives the necessary and sufficient conditions for girth larger than 12. 
\begin{theorem}\label{thm:girth>12-for2} 
 $\girth(H_2) >12$ if and only if $C_{21}C_{12}^\tr C_{21}\triangle C_{21}=0$. 
\end{theorem} 


\begin{example} 
 The $2N\times 5N$ matrix $H_2$ with $P_2, P_3, P_4,  P_5,$ and  $C_{21}$ listed below (in this order) has girth 16 and, equivalently,  the  $3N\times 3N$ matrix $ C_{21}$ has girth 8:
$$  \begin{bmatrix}
0&0&\matr{x} \\   
1& 0&0\\  
0&x^7& 0\end{bmatrix}, 
 \begin{bmatrix}
   0&0&\matr{x^3}  \\ 
       x^5&0&0         \\ 
          0&x^{11}&0      \end{bmatrix}, 
  \begin{bmatrix}
\matr{x^{6}}&0&0 \\ 
     0&x^{23}& 0\\ 
           0&0&x^{29}         \end{bmatrix},   \begin{bmatrix}
  0&x^{15}&0 \\ 
     0&0&\matr{x^{19}}\\ 
x^{42}&0&0\end{bmatrix}, \begin{bmatrix}
x^6+1&x^{15}&\matr{x} +x^3\\   
1+x^5& x^{23}+1&x^{19}\\  
x^{42}&x^7+x^{11}& x^{29}+1\end{bmatrix}.$$ 
The pre-lifted protograph used for  $[P_1 \ P_2 \ P_3\ P_4 \ P_5]$ 
corresponds  to $\begin{bmatrix}  1&x&x&1&x^2\end{bmatrix}. $\end{example}



\begin{example} 
In this example, we  start from a  $5\times 5 $ pre-lifted protograph $\begin{bmatrix}  1& x&x^2&x^3&x^4\end{bmatrix} $ for sub-matrix $[P_1 \ P_2 \ P_3\ P_4 \ P_5]$ because it has  girth 6. 
  This would insure that the structures that limit the girth to 12, 14, and 16 (listed in~\cite{4276926}) are all avoided. The lifted matrices $[P_2 \ P_3\ P_4 \ P_5]$ are (in order) 
%
 {\small $$\begin{bmatrix} 
   0&0&0&  0&x \\
  1& 0&0&0&0   \\  
 0&x^0& 0&0&0  \\
   0&0&x^{51}&0&0 \\    
0&0&0&x^{68}& 0\end{bmatrix},
\begin{bmatrix} 
        0 &0&0&x^2&0\\           0 &0&0& 0&1\\
                   x^{12} & 0&0&0&0 \\            0&x^{79}&0 &0& 0\\            0&0& x^{94}&0& 0\end{bmatrix}\begin{bmatrix} 
          0&0&x^4&0&0 \\        0&0&0& x^4&0\\          0&0&0& 0& x^{26} \\   x^{109}&0& 0&0&0   \\  0&x^{180} &0& 0&0\end{bmatrix}, \begin{bmatrix} 
0&x^9&0&0&0   \\ 
   0&0&x^{10}&0&0  \\ 
 0&0&0&x^{67}&0   \\
 0& 0&0&0&x^{164}  \\
  x^{309}&0&0&0&0 \end{bmatrix},$$}
giving a matrix $H_2$ of girth  20 for $N=458$ and, equivalently, the matrix
$$C_{21}=P_1+P_2+\cdots+P_5= \begin{bmatrix}1&x^9&x^4&x^2&x\\1&1&x^{10}&x^4&1\\x^{12}&1&1&x^{67}&x^{26}\\
x^{109}&x^{79}&x^{51}&1&x^{164}\\ x^{309}&x^{180}&x^{94}&x^{68}&1
\end{bmatrix} $$
of girth 10 for $N=458.$ 
\end{example}

\begin{remark} Note that by taking  a $3\times 5$ submatrix   of $P_1+P_2+\cdots+P_5$ above, we obtain a $(3,5)$-regular LDPC code of girth 10. 
 For example, the matrix $H_{\rm sub}$ below gives a code of minimum distance 24 and  girth 10 for a minimum value $N=101$ (and other larger) $$H_{\rm sub} \defeq   \begin{bmatrix}1&x^9&x^4&x^2&x\\1&1&x^{10}&x^4&1\\x^{12}&1&1&x^{67}&x^{26}\end{bmatrix} .
$$

Since  all protograph-based $n_cN\times n_vN$ parity-check matrices  can be seen as a submatrices of  $n_vN\times n_vN$ square matrices that are equal to  sums of $n_v$ lifted permutation matrices,  we can construct  codes by constructing sums of permutation matrices with good girth, and either use them as the parity-check matrices (for a proper value of $N$), or take submatrices,  like we did in the above example. 
 \end{remark} 


  
\section{The girth of  $n_cN\times n_vN$ matrices $H_{n_c}$, $n_c=3,4$} \label{sec:3bynv}

This section provides minimal necessary and sufficient conditions and associated algorithms to find $3\times n_v$ and $4\times n_v$ 
QC-LDPC protograph-based codes of various girth $2m\geq 6$. They start by constructing a $2\times n_v$ protograph-based code of girth $\geq 2m$ and expanding it to a $3\times n_v$, then to $4\times n_v.$ 
Cases of $n_c\geq 5$ can be solved similarly.  

As before, we will consider $H_{n_c}$ to be in the {\em  reduced form}, i.e., it  has identity matrices on the first row and first column,  since the conditions are simpler to see.  In addition, we will assume that 
%
the matrix  $H_{n_c}$   is composed of circulant matrices $ x^{i_{l}}, x^{j_{l}}, x^{k_{l}}$, for all $l\in[n_v]$, with $i_1=j_1=k_1=0$. 
  \subsection{Case of  $4<\girth(H_{n_c})\leq 12$} \label{sec:3byn-girth}
  In this section we will consider $H_{n_c}$ as defined below, for $n_c=2,3,$ and $4$, respectively,  alongside the corresponding matrices $C_{H_{n_c}}$ and $\matr{C_{ij}}$. As mentioned above, we assume, without loss of generality, that $i_1=j_1=k_1=0$ (i.e., $H_{n_c}$ is in reduced form): 
  \begin{align}\label{H-circulant_2}& \matr{H_{2}}=\begin{bmatrix} 1&\cdots&1\\ x^{i_{1}} &\cdots&x^{i_{n_v}}\end{bmatrix}, \quad C_{H_2}\defeq  \begin{bmatrix} 0&C_{12} \\ C_{21}&0\end{bmatrix}; \\\label{H-circulant_3}& 
  \matr{H_{3}}=\begin{bmatrix}\matr{H_{2}} \\\begin{matrix} x^{j_{1}} &\cdots&x^{j_{n_v}}\end{matrix}\end{bmatrix}=\begin{bmatrix}1&\cdots&1\\ x^{i_{1}} &\cdots&x^{i_{n_v}}\\ x^{j_{1}}&\cdots&x^{j_{n_v}}\end{bmatrix}, C_{H_3}\defeq  \begin{bmatrix} 0&C_{12} &C_{13}\\ C_{21}&0&C_{23}\\ C_{31}&C_{32} &0\end{bmatrix}= \begin{bmatrix} C_{H_2} &C_{3}\\ C_{3}^\tr & 0\end{bmatrix}; \\
\label{H-circulant_4}&  \matr{H_{4}}=\begin{bmatrix}\matr{H_{3}} \\\begin{matrix} x^{k_{1}} &\cdots&x^{k_{n_v}}\end{matrix}\end{bmatrix}= \begin{bmatrix} 1&\cdots&1\\ x^{i_{1}} &\cdots&x^{i_{n_v}}\\ x^{j_{1}}&\cdots&x^{j_{n_v}}\\ x^{k_{1}} &\cdots&x^{k_{n_v}}\end{bmatrix},  C_{H_4}\defeq  \begin{bmatrix} 0&C_{12} &C_{13}&C_{14}\\ C_{21}&0&C_{23}&C_{24}\\ C_{31}&C_{32} &0&C_{34} \\C_{41}&C_{42}&C_{43}&0\end{bmatrix} = \begin{bmatrix} C_{H_3} &C_{4}\\ C_{4}^\tr & 0\end{bmatrix};
  \\\label{C_ij, C_j}& \begin{matrix}\matr{C_{12}}=C_{21}^\tr \defeq &\sum\limits_{l=1}^{n_v} x^{-i_{l}}, & \matr{C_{23}}=C_{32}^\tr\defeq& \sum\limits_{l=1}^{n_v} x^{i_{l}-j_{l}} \\  \matr{C_{13}}=C_{31}^\tr \defeq &\sum\limits_{l=1}^{n_v} x^{-j_{l}}, & \matr{C_{24}}=C_{42}^\tr\defeq& \sum\limits_{l=1}^{n_v} x^{i_{l}-k_{l}} \\  \matr{C_{14}}=C_{41}^\tr \defeq &\sum\limits_{l=1}^{n_v} x^{-k_{l}}, & \matr{C_{34}}=C_{43}^\tr\defeq& \sum\limits_{l=1}^{n_v} x^{j_{l}-k_{l}} \end{matrix}, \quad C_{3}\defeq \begin{bmatrix} C_{13}\\C_{23} \end{bmatrix}, \quad C_{4}\defeq \begin{bmatrix} C_{14}\\C_{24}\\C_{34} \end{bmatrix}.\end{align}


\begin{theorem}\label{thm:girth6-4}  Let $H_4$ and $C_{H_4}$ be defined as in \eqref{H-circulant_4} and \eqref{C_ij, C_j}. 
Then $\girth(H_4) >4$ if and only if 
each one of the six  sets  
$\{i_1, \ldots, i_{n_v}\},   \{j_1,\ldots, j_{n_v}\},  \{k_1,  \ldots, k_{n_v}\}, \{i_1-j_1, \ldots, i_{n_v}-j_{n_v}\},\\\{i_1-k_1, \ldots, i_{n_v}-k_{n_v}\}, \text{ and } \{j_1-k_1,  \ldots, j_{n_v}-k_{n_v}\}$ 
is of maximal size.
\end{theorem} 
\begin{IEEEproof}
In order to avoid 4-cycles, we need to insure that  the codes based on the $2\times n_v$ sub-protographs have no 4 cycles, for all such sub-protographs, and, equivalently, that     $\matr{C_{ij}}\triangle 0=0$, for all $1\leq i <j\leq 4$.  The claim  on the  sets above follows from this.   \end{IEEEproof} 

\begin{remark} \label{thm:girth6-3}The conditions for $\girth(H_3) >4$ are obtained from Theorem~\ref{thm:girth6-4} by ignoring the sets above that contain any $k_j$, i.e., $\girth(H_3) >4$ if and only if 
each one of the 3  sets  
$\{i_1, \ldots, i_{n_v}\},   \{j_1,\ldots, j_{n_v}\},  \{i_1-j_1, \ldots, i_{n_v}-j_{n_v}\}$ is of maximal size.
\end{remark} 

 We now present an algorithm to  choose these exponents in which we first choose  $i_1, i_2, \ldots , i_{n_v}$ (sequentially), and then choose $j_1, j_2, \ldots, j_{n_v}$, and then all $k_1, k_2, \ldots, k_{n_v}$, such that, at each step, the conditions of Theorem~\ref{thm:girth6-4} are satisfied. This algorithm is an extension of Algorithm $A_{2,g>4}$ for $2N\times n_vN$ parity-check matrices $H_2$. To obtain an algorithm $A_{3,g>4}$ for $n_c=3$, we simply  ignore Step 4 that chooses the exponents $k_l$, $l\in [n_v]$.

\noindent \fbox{\bf Algorithm $A_{4,g>4}$}\\
  Step 1:  Set  $i_1= j_1=k_1=0$.\\
    Step 2: Set  $l=1$ and repeat until $l=n_v$:   Let $l:=l+1$; choose $i_{l} \notin \{i_a \mid a\in [l-1]\}$. \\
     Step 3: Set  $l=1$ and repeat until $l=n_v$: Let $l:=l+1$; choose  $j_{l} \notin \{j_a,  i_l+ (j_a-i_a) \mid a\in [l-1]\}.$ \\
    Step 4: Set  $l=1$ and repeat until $l=n_v$: Let $l:=l+1$; choose $k_{l} \notin \{k_a,  i_l+ (k_a-i_a),  j_l+ (k_a-j_a)\mid a\in [l-1]\}.$ 

We now extend to larger girth.
\begin{theorem}\label{thm:girth8-for4} Let $H_4$ and $C_{H_4}$ be defined as in \eqref{H-circulant_4} and \eqref{C_ij, C_j}. 
Then $\girth(H) >6$ if and only if, for all $l\in[n_v]$, each one of the  sets 
$$\{i_l-i_s,  j_l-j_s,  k_l-k_s \mid s\in[n_v], s\neq l \},  \{i_s, i_s-j_s+j_l, i_s-k_s+k_l\mid s\in[n_v], s\neq l \},  $$
$$ \{j_s, j_s-i_s+i_l, j_s-k_s+k_l\mid s\in[n_v], s\neq l\}, \{k_s, k_s-i_s+i_l,k_s-j_s+j_l\mid s\in[n_v], s\neq l \}$$
is of maximal size. 

\end{theorem} 
\begin{IEEEproof}  From Theorem~\ref{mbyn-girth>6} we see that in order to avoid 6-cycles, we need to insure  that, for all $ l\in [n_v]$,  and all $s,t\in [n_v]\setminus \{ l\} $, 
 $$\left\{\begin{matrix} 
 (C_{12}x^{i_l}+ C_{13}x^{j_l}+ C_{14}x^{k_l})\triangle 1 =0\\
 (C_{21}+ C_{23}x^{j_l}+ C_{24}x^{k_l})\triangle x^{i_l} =0 \\
(C_{31}+ C_{32}x^{i_l}+ C_{34}x^{k_l})\triangle x^{j_l} =0\\
(C_{41}+ C_{42}x^{i_l}+ C_{43}x^{j_l})\triangle x^{k_l} =0
\end{matrix}\right.  \Longleftrightarrow  \left\{\begin{matrix} 
 \sum\limits_{{s=1}\atop {s\neq l}}^{n_v} \left(x^{i_l-i_s}+x^{j_l-j_s} +x^{k_l-k_s} \right)\triangle 1 =0\\
 \sum\limits_{{s=1}\atop {s\neq l}}^{n_v} \left(x^{i_s}+x^{i_s-j_s+j_l} +x^{i_s-k_s+k_l} \right)\triangle x^{i_l} =0\\
 \sum\limits_{{s=1}\atop {s\neq l}}^{n_v} \left(x^{j_s}+x^{j_s-i_s+i_l} +x^{j_s-k_s+k_l}\right)\triangle x^{j_l} =0\\
 \sum\limits_{{s=1}\atop {s\neq l}}^{n_v} \left(x^{k_s}+x^{k_s-i_s+i_l} +x^{k_s-j_s+j_l}\right)\triangle x^{k_l} =0
 \end{matrix}\right., $$

\noindent from which the claim follows. \end{IEEEproof}

\noindent \fbox{\bf Algorithm $A_{4,g>6}$}\\
  Step 1:  Set  $i_1=j_1=k_1=0$. \\
  Step 2:  Set  $l=1$ and repeat until $l=n_v$:   Let $l:=l+1$; choose $i_{l} \notin \{i_a \mid a\in [l-1]\}$. \\
 Step 3: Set  $l=1$ and repeat until $l=n_v$: Let $l:=l+1$; choose  $j_{l} \notin \{j_a, i_t+ j_a-i_a, i_l+j_a-i_t \mid a\in [l-1], t\in [n_v]\}.$ \\
    Step 4: Set  $l=1$ and repeat until $l=n_v$: Let $l:=l+1$; choose   
$k_l\notin \{i_l+(k_s-i_t), j_l+(k_s-j_t), i_t+(k_s-i_s), j_t+(k_s-j_s), j_l+(k_s-i_s)+(i_t-j_t), i_l+(k_s-j_s)+(j_t-i_t)\mid s\in[l-1], t\in[n_v]\}$


\begin{example} \label{N-values-4} We construct two  $4\times 8$ protograph-based matrices, one of girth larger than 4, and one of girth larger than 6,  starting from the  $2\times 8$ matrix $H_{2,g>4}$ from  Example~\ref{example-2byn} and adding a 3rd and a 4th row such that the conditions of the Algorithms $A_{4,g>4}$ and $A_{4,g>6}$ are satisfied. Choosing the exponent at each step as the smallest positive integer not in the forbidden set yields
$$H_{4,g>4}= \begin{bmatrix} 1&1&1&1&1&1&1&1\\ 1& x&x^2& x^3&x^{4}& x^5 &x^6&x^7 \\1&x^{2}& x&x^5&x^7 &  x^{3}&x^{10}&x^{4} \\1&x^{3}&x^{5}&x& x^{9}&x^{2}&x^{7}&x^{11}
\end{bmatrix}, H_{4,g>6}= \begin{bmatrix} 1&1&1&1&1&1&1&1\\ 1& x&x^2& x^3&x^{4}& x^5 &x^6&x^7 \\1&x^{8}& x^{15}&x^{21}&x^{26} &  x^{32}&x^{39}&x^{47} \\1&x^{9}&x^{17}&x^{24}& x^{30}&x^{37}&x^{45}&x^{54}
\end{bmatrix}.$$
 
 \end{example}
 \begin{remark}  At each step, we  can choose $i_{l} $, $j_l$, and $k_l$ to be the minimum positive integer such that they satisfy the conditions of the algorithm (as done in Example \ref{N-values-4}), but the resulting matrix may not necessarily have the smallest $N$ possible for that girth, 
 nor be the best code. 
The above algorithms can be modified to, e.g., select exponents randomly, avoiding those values in the forbidden set, or so that they are larger than the maximum value in the forbidden set. Different choice of exponents will yield different minimum $N$ (see also Remark \ref{smallestNg10}). Finally, we note that the algorithms can also be modified to pick exponents column-by-column, rather than row-by-row as presented here; see \cite{sm21} for examples. 
 \end{remark} 

Similar theorems can be stated for $\girth(H_4)>8$ and  $\girth(H_4)>10$, but the number of conditions increase; so, for clarity we only present them  for $n_c=3$ and refer the reader to Remark~\ref{extendto4} for how to extend these conditions to the case $n_c=4$. We also refer the reader to~\cite{gomezfonseca2021necessary}, where a different approach was taken to list these conditions.

\begin{theorem}\label{thm:girth10-3} Let $H_3$ and $C_{H_3}$ be defined as in \eqref{H-circulant_3} and \eqref{C_ij, C_j}. Then $\girth(H_3) >8$ if and only if  each two  of  the following sets of differences 
$$\{i_u-i_v\mid u\neq v, u,v\in[n_v]\} , \{j_u-j_v\mid u\neq v, u,v\in[n_v]\}, \{(i_u-j_u)-(i_v-j_v)\mid u\neq v, u,v\in[n_v]\}$$
contains non-equal values and each set is of maximal size. 

Equivalently, $\girth(H_3) >8$  if and only if each one of the three sets 
$\{i_u-i_v, j_u-j_v\mid u\neq v, u,v\in[n_v]\},\{i_u-i_v, (i_u-j_u)-(i_v-j_v)\mid u\neq v, u,v\in[n_v]\} ,\{j_u-j_v, (i_u-j_u)-(i_v-j_v) \mid u\neq v, u,v\in[n_v]\}$
is of maximal size. 
\end{theorem}   
\begin{IEEEproof} We have  $\girth(H_3) >8$ if and only if $\girth(C_{H_3})=6$ if and only if $C_{H_3}^2\triangle I=0$.  By expanding this last equality  into the equivalent conditions we obtain:  
 \begin{align*}
& \left\{ \begin{matrix} 
(C_{12}C_{21}+ C_{13}C_{31})\triangle I=0\\
(C_{21}C_{12}+ C_{23}C_{32})\triangle I=0\\
(C_{31}C_{13}+ C_{32}C_{23})\triangle I=0 \end{matrix}\right. \Longleftrightarrow  \left\{ \begin{matrix} 
\sum\limits_{u,v\in[n_v]}x^{i_u-i_v}+ \sum\limits_{u,v\in[n_v]}x^{j_u-j_v}&\triangle 1=0\\
\sum\limits_{u,v\in[n_v]}x^{i_u-i_v}+ \sum\limits_{u,v\in[n_v]}x^{(i_u-j_u)-(i_v-j_v)}&\triangle 1=0\\
\sum\limits_{u,v\in[n_v]}x^{j_u-j_v}+ \sum\limits_{u,v\in[n_v]}x^{(i_u-j_u)-(i_v-j_v)}&\triangle 1=0 \end{matrix}\right. \end{align*}
The claim follows.\end{IEEEproof} 

\begin{remark} Since $C_{H_3}$ has  girth 6, it means that $C_{ij}$ have girth 6, for all $1\leq i<j\leq 3$. Therefore,  all  $2N\times n_vN$ sub-matrices have girth 12 when, overall, $H_3$ has girth 10. 
\end{remark} 

%
 \noindent \fbox{\bf Algorithm $A_{3,g>8}$} \\
Step 1:  Set  $i_1=j_1=0$. \\
Step 2: Set  $l=1$ and repeat until $l=n_v$:   Let $l:=l+1$; choose $i_l \notin \{ i_u+ i_s-i_t \mid  u, t, s \in [l-1]\}.$ \\
%
Step 3: Set  $l=1$ and repeat until $l=n_v$:   Let $l:=l+1$; choose $j_l \notin \{j_u+j_s-j_t,   j_{u} + i_a-i_b, j_u+(j_s-i_s)-(j_t-i_t), i_l+ i_a-i_b+(j_{u} -i_u), i_l+(j_{u} -i_u) +(j_{s} -i_s)- (j_{t} -i_t), i_l+ j_s-j_t +(j_{u} -i_u)  \mid \ a, b \in[n_v], u, s,t\in [l-1]\} $.\footnote{We note that Step 3 in Algorithm $A_{3,g>8}$ (and also in $A_{3,g>10}$ later), the range $t\in[l-1]$ assumes that the exponents $j_l$ are chosen to be increasing in value with $l$. This is how we have implemented the algorithms in our examples. However, for general exponent selection (increasing in value or not) we should amend Step 3 to have $t\in[l]$ to ensure that the conditions in the corresponding theorem are met.\label{footnotetl}}



 \begin{remark} \label{smallestNg10}
Given a parity-check matrix $H_{3}$ that meets the conditions of Theorem \ref{thm:girth10-3} (i.e., it can achieve girth larger than 8) then the lifting factors $N\in \mathbb{Z}^+$ for which the parity-check matrix has girth 10 (or larger) are given by
 
\noindent $N\nmid  \left\{i_s+i_t-i_u-i_v,  j_u-j_v +i_s-i_t, j_u-j_v+ j_s-j_t, j_u-j_v +(j_s-i_s)-(j_t-i_t), \right.\\\left. (j_s-i_s)-(j_t-i_t)+ (j_u-i_u)-(j_v-i_v), (j_s-i_s)-(j_t-i_t) +i_u-i_v\mid s,t,u,v\in[n_v]\}\right.,$ 

\noindent 
 where the smallest such value is denoted $N_\textrm{min}$ and the notation $a\nmid b$ denotes that $a$ does not divide $b$. Similar statements can be made to determine viable (and minimum) $N$ for general $H_{n_c}$ and desired girth by using the conditions in the associated theorem or, equivalently, algorithm. We note picking a larger $N$ than $N_\textrm{min}$ may often yield better performance provided that the girth is maintained, see Section \ref{sec:simulations}.
\end{remark} 
\begin{example} \label{girth10-Example} Note that in Example~\ref{example-2byn},  we used Algorithm $A_{2,g>8}$ (the first part of Algorithm $A_{3,g>8}$) 
to obtain a $2N\times n_vN$ matrix $H_{2,g>8}$  of girth 12 for $N=77$. The circulants $x^{i_l}$ are   
%
 $\begin{bmatrix} x^{i_1}&x^{i_2}&x^{i_3}&\cdots&x^{i_8}\end{bmatrix} = \begin{bmatrix}1& x&x^3&x^7&x^{12}&x^{20}&x^{30}&x^{44}\end{bmatrix}.  $ The matrix  $C_{21} =  1+ x+x^3+x^7+x^{12}+x^{20}+x^{30}+x^{44}$  has, equivalently, girth 6 for $N=77.$ 

 Therefore, we need to choose the row $ \begin{bmatrix} x^{j_1}&x^{j_2}&x^{j_3}&\cdots&x^{j_8} \end{bmatrix} $ such that $C_{31}= x^{j_1}+x^{j_2}+x^{j_3}+\cdots+x^{j_8} $  has girth 6, but also such that the resulting matrix $C_{32}=x^{j_1-i_1}+x^{j_2-i_2}+x^{j_3-i_3}+\cdots+x^{j_8-i_8}  $ has girth 6, and  also $C_H$ has girth 6.  
So we choose  $j_l$ such that  the difference between $j_l$ and any exponent already found in $C_{21}$ or $C_{31}$
 does not appear among the differences of the exponents already found,  and we impose the same for $j_l-i_l$, i.e., 
that  the difference between $(j_l-i_l)$ and any exponent already found in $C_{21}$ or $C_{32}$
 does not appear among the differences of the exponents already found.

We obtained the following matrix with Tanner graph of girth 10 for $N_\textrm{min}=514$ (applying Remark \ref{smallestNg10})
 $$H=\begin{bmatrix} 1&1&1&1&1&1&1& 1\\1& x&x^3&x^7&x^{12}&x^{20}&x^{30}&x^{44}\\ 1& x^{66} &x^{461}& x^{106}& x^{144}& x^{194}& x^{274} &x^{385}\end{bmatrix}.$$    

\noindent Note that $C_{12}$, $C_{13}$, and $C_{23}$ all have girth 6, giving 3 $(2,8)$-regular codes of girth 12. 
 \end{example}
   

  The following lemma extends Lemma~\ref{2byn-girth12-construction} for $2N\times n_vN$ parity-check matrices $H_2$ of $\girth(H_2)> 8$ to $3N\times n_vN$ matrices $H_3$. It gives an easy way to choose the next exponent values  such that they are larger than the ones in the forbidden sets, i.e., sets that would decrease the girth to 8 or lower.   
 \begin{lemma} \label{girth10-construction} Let $H_3$ and $C_{H_3}$ be defined as in \eqref{H-circulant_3} and \eqref{C_ij, C_j}.  Let $i_l$ and $j_l$ be defined recursively as
\begin{align*} &\left\{ \begin{matrix}  i_1= 0, \\ i_l=1+2i_{l-1},\>\> l\geq 2,\end{matrix} \right.\text{ and } \left\{ \begin{matrix}  j_1= 0, \>\>j_2=1+i_2+2i_{n_v},\\ j_l=1+ 2j_{l-1} +i_{l},\>\>  
l\geq 3. \end{matrix} \right.\end{align*}

\noindent  The Tanner graph of the code with parity-check  matrix $H_3$ has girth 10 for  some $N$.  \end{lemma} 

\begin{remark} Instead of the choice of $j_l$ above, we can alternatively  choose, for example,  
$j_l= 1+3j_{l-1}$ or $j_l=1+j_{l-1}+\max \{j_{l-2}, i_{n_v}\} 
+\max\{j_{l-3}, i_{n_v}\}$ to obtain a  QC-LDPC code  matrix $H_3$ with girth 10 for  some $N$. Such choices hold since each of the  chosen values satisfy the conditions of Algorithm $A_{3,g>8}$, where they can be seen to be, at each step,  larger than the largest forbidden value. \end{remark}

We exemplify this easy method below. 
\begin{example} We construct a $3\times 7$ matrix based on the Lemma~\ref{girth10-construction} to obtain the first matrix $H_{3,g>8}$ below of girth 10 for $N_\textrm{min}=433$. After reducing the exponents modulo $N=433$, this matrix  is equal to the second matrix,  which has girth 10 for $N_\textrm{min}=347$:   
$$H_{3,g>8}=\begin{bmatrix} 
1&1&1&1&1&1&1\\
1& x           &x^3       &x^7       &x^{15}  &x^{31}   &x^{63}\\ 
1& x^{128} &x^{260}& x^{528}& x^{1072}& x^{2176}& x^{4416} \end{bmatrix}\equiv 
\begin{bmatrix} 
1&1&1&1&1&1&1\\
1& x           &x^3       &x^7       &x^{15}  &x^{31}   &x^{63}\\ 
1& x^{128} &x^{260}& x^{95}& x^{206}& x^{11}& x^{86} \end{bmatrix}.$$    
 If we write $260=-87$ and $206=-141$, we obtain the first matrix below of girth 10 for  the new minimum value $N=327$, for which  $-141=186$ and $-87=240$, as shown in  the second matrix below
 $$H_{3,g>8}=\begin{bmatrix} 
1&1&1&1&1&1&1\\
1& x           &x^3       &x^7       &x^{15}  &x^{31}   &x^{63}\\ 
1& x^{128} &x^{-87}& x^{95}& x^{-141}& x^{11}& x^{86} \end{bmatrix}\equiv
  \begin{bmatrix} 
1&1&1&1&1&1&1\\
1& x           &x^3       &x^7       &x^{15}  &x^{31}   &x^{63}\\ 
1& x^{128} &x^{240}& x^{95}& x^{186}& x^{11}& x^{86}  \end{bmatrix}. $$   
This last matrix has girth 10 for $N_\textrm{min}=278$. 

Note that this $N$ is not the minimum for which a code can be found with girth 10, since the minimum with the algorithm is $N=219$. But it can be easily  obtained by hand. 
\end{example} 
 

\begin{theorem}\label{thm:girth>10}  Let $H_3$ and $C_{H_3}$ be as defined in \eqref{H-circulant_3} and \eqref{C_ij, C_j}.  Then $\girth(H) >10$ if and only if,   for all $l\in [n_v]$, 
\begin{enumerate} 

\item each two  of  the four sets of differences 

$\{i_u-i_v\mid u\neq v, u,v\in[n_v], u\neq l\} ,  \{j_u-j_v\mid u\neq v, u,v\in[n_v], u\neq l\}, \\
 \{-j_u+ j_v-i_v+i_l\mid u\neq v, u,v\in[n_v], v\neq l\},\{-i_u+i_v-j_v+j_l\mid u\neq v, u,v\in[n_v], v\neq l\} $
 
contain non-equal values, for all $l\in [n_v]$, and each set is of maximal size.
\item  each two  of  the four  sets of differences 

$ \{i_u-j_u+j_v \mid u\neq v, u,v\in[n_v], v\neq l\} ,  \{i_u-i_v+i_l\mid u\neq v, u,v\in[n_v], v\neq l\}, \\
 \{(i_u-j_u)-(i_v-j_v)+i_l\mid u\neq v, u,v\in[n_v], v\neq l\}, \{i_u-j_v+j_l\mid u\neq v, u,v\in[n_v], v\neq l\}$
contain non-equal values, and each set is of maximal size.
\item each two  of  the four sets of differences 

$\{j_u-i_u+i_v\mid u\neq v, u,v\in[n_v], v\neq l\} ,  \{j_u-i_v+i_l\mid u\neq v, u,v\in[n_v], v\neq l\}, \\
 \{j_u- j_v+j_l\mid u\neq v, u,v\in[n_v], v\neq l\},\{j_u-i_u+i_v-j_v+j_l\mid u\neq v, u,v\in[n_v], v\neq l\} $
contain non-equal values, and each set is of maximal size.
\end{enumerate} 
\end{theorem} 
\begin{IEEEproof} We apply the condition $C_{H_3}^2H_3\triangle (H_3+ C_{H_3}H_3)=0$
to obtain 
$$\left\{\begin{matrix} 
\left(C_{12}C_{21} +C_{13}C_{31} +C_{13}C_{32}x^{i_l} +C_{12}C_{23}x^{j_l}  \right)\triangle   (C_{12}x^{i_l} +C_{13}x^{j_l} +I) =0,\\ 
\left(C_{23}C_{31} +\left(C_{21}C_{12} +C_{23}C_{32}\right)x^{i_l} +C_{21}C_{13}x^{j_l} \right) \triangle  (C_{21}+ C_{23}x^{j_l}+x^{i_l}) =0,\\ 
\left(C_{32}C_{21} +C_{31}C_{12}x^{i_l} +\left(C_{31}C_{13}+C_{32}C_{23}\right) x^{j_l} \right) \triangle  (C_{31}+ C_{32}x^{i_l}+x^{j_l}) =0,
\end{matrix}\right. \Longleftrightarrow$$
{\small \begin{align*} &\sum\limits_{u,v\in[n_v]}\left(x^{i_u-i_v}+x^{j_u-j_v} + x^{i_l+(j_u-i_u)- j_v} +x^{j_l+(i_u-j_u)- i_v}\right) \triangle \left(1+ \sum\limits_{u\in[n_v]}\left(x^{i_l-i_u} + x^{j_l-j_u} \right)\right)=0, \\
&\sum\limits_{u,v\in[n_v]}\left(x^{(i_u-j_u) +j_v}+x^{i_l+i_u-i_v} +x^{i_l+(i_u-j_u)+(j_v-i_v)}+ x^{j_l+(i_u-j_v)} \right)\triangle  \left(x^{i_l}+ \sum\limits_{u\in[n_v]}\left(x^{i_u} + x^{j_l+i_u-j_u}\right)\right)=0,\\
&\sum\limits_{u,v\in[n_v]}\left(x^{(j_u-i_u) +i_v}+x^{i_l+j_u-i_v} +x^{j_l+(j_u-j_v)} +x^{j_l+(j_u-i_u) +(i_v-j_v)}\right) \triangle  \left(x^{j_l}+ \sum\limits_{u\in[n_v]}\left(x^{j_u} + x^{i_l+j_u-i_u} \right)\right)=0 .\end{align*}
}
 
\noindent The three equalities above hold if any two monomials on the left side of the triangle operator are not equal, unless they are equal to one of the monomial on the right side of the triangle operator. We obtain the claim of the theorem. 
\end{IEEEproof} 

The following algorithm $A_{3, g>10}$ uses Theorem \ref{thm:girth>10} to construct a parity-check matrix $H_3$ such that the girth is 12. Similar to the above, the exponents $i_u$, $u\in [n_v]$ are chosen first, i.e., it chooses the row of the matrices $x^{i_l}$ such that the girth of the $2N\times n_vN$ matrix is equal to 12. (Note that this matrix would be the same as for $\girth(H_2)>8$ from $A_{2, g>8}$.) Following this, one more row is added with the additional conditions above to insure that the girth is 12 rather than $\geq 10$ as in $A_{3, g>8}$. 


\noindent \fbox{\bf Algorithm $A_{3, g>10}$}\\
Step 1:  Set  $i_1=j_1=0$. \\
Step 2: Set  $l=1$ and repeat until $l=n_v$:   Let $l:=l+1$; choose $i_l \notin \{ i_u+ i_s-i_t \mid  u, t, s \in [l-1]\}.$ \\%
%
Step 3: Run loop until $l=n_v$:   Let $l:=l+1$; choose $j_l \notin \{ i_a-i_b+ j_{s}, 
i_a+j_s-j_t+(j_u-i_u), 
i_a+i_b-i_c+(j_s-i_s), 
-i_a+j_s+j_u-(j_t-i_t), 
j_s-(j_t-i_t)+(j_u-i_u),
i_a+(j_s-i_s)- (j_t-i_t)+(j_u-i_u), 
 i_l+ i_{a} -j_t+ (j_s-i_s)+(j_u-i_u),
 i_l-i_a+ j_s - (j_t-i_t)+(j_u-i_u),
 \\i_l-i_a+ j_s+j_u-j_t,
 i_l+i_a-i_b-i_c + j_s, 
 j_s+j_u-j_t, 
 \mid   a, b, c \in [n_v], s, u, v \in[l-1], t\in [l]\} $.\footnote{Again, for general exponent selection (increasing in value or not) we should amend Step 3 to have $t\in[l]$ to ensure that the conditions in the corresponding theorem are met (see Footnote \ref{footnotetl}).}

 The following is such an example.  We start from $H_{2,g>8}$ and use the algorithm $A_{3, g>10}$ to find the third row. 
  \begin{example} 
 The  matrix 
  $$\matr{H}=\begin{bmatrix}  1&1&1&1&1&1&1& 1\\1& x&x^3&x^7&x^{12}&x^{20}&x^{30}&x^{44}\\ 1& x^{66} &x^{144}&  x^{232}& x^{336}& x^{526}& x^{664}& x^{747}\end{bmatrix}$$
 has girth 12 for $N_\textrm{min}=1245$ (length $n_vN_\textrm{min}=9960$), for example. 
 \end{example}

\begin{remark} An alternative way to obtain a girth 12 matrix is by starting from a girth 10 matrix and modifying the exponent of a circulant matrix that is a component of a 10-cycle. Since a girth 10 matrix $H_3$ must have all $2 \times n_v$ and $3\times 2$  sub-protographs lifted to girth 12 codes, we can check the girth of the $3\times 3$, $3\times 4$, etc., matrices, to find out which ones are of girth 12 (if any) and which are of girth 10. If we find out that an entry decreases the girth from 12 to 10, we can change its exponent to a much larger one to break the 10 cycle. We do this in the following example. \end{remark} 
\begin{example} \label{girth12-from 10} Let us consider the $(3,5)$-regular submatrix $\matr{H}_{3,g>8}$ below obtained from the $(3,8)$-regular QC-LDPC code of girth 10 for $N=514$ constructed  in Example~\ref{girth10-Example}. Note that, for this $N$, all $2\times 3$ and $2\times 5$ submatrices of the $3\times 5$ matrix $H_{3,g>8}$  correspond to $(2,3)$-regular and $(2,5)$-regular protograph-based codes, respectively,  with girth 12, since $C_{H_{3,g>8}}$ has girth 6 and hence the submatrices $C_{ij}$ all have girth 6 (they cannot be higher), resulting in associated matrices of girth 12. Since the girth of $H_{3,g>8}$ is 10, there must be some 10-cycles in $H_{3,g>8}$ in a $3\times 3$ submatrix of $H_{3,g>8}$. We check the girth of each $3\times 4$ submatrix  and find  that the $3\times 4$ submatrix  obtained from columns 1, 2, 3, and 5 of $H_{3,g>8}$ has girth 12, and so does the $3\times 5$ matrix obtained from $H_{3,g>8}$ by masking  $x^{144}$ (substituting it with 0). Hence the 10-cycle visits this  circulant. We make a substitution, for example, $x^{244}$  instead of $x^{144}$, to obtain $H_{3,g>10}$ below of girth 12 for $N_\textrm{min}=328$:
$$\matr{H}_{3,g>8}=\begin{bmatrix} I&I&I&I&I\\1& x&x^7&x^{12}&x^{20}\\ 1& x^{66} & x^{106}& x^{144}& x^{194}\end{bmatrix} \rightsquigarrow\matr{H}_{3,g>10}=\begin{bmatrix} I&I&I&I&I\\1& x&x^7&x^{12}&x^{20}\\ 1& x^{66} & x^{106}& x^{244}& x^{194}\end{bmatrix}.$$
We note that $H_{3,g>8}$ has girth 10 for $N_\textrm{min}=158$ and $H_{3,g>10}$ has girth 10 for $N_\textrm{min}=222$, obtained using Remark \ref{smallestNg10}, but $H_{3,g>8}$ cannot achieve girth 12 for any $N$.
\end{example}
In Appendix~\ref{pre-lift-observed-2},  we revisit Example~\ref{girth12-from 10} to show how we can use the girth 10 construction together with the pre-lifting techniques presented in Section~\ref{sec:girth>12}, in order to obtain a girth 12 code and possibly increase the minimum distance.    

We conclude this section with a remark concerning the extension of the $n_c=3$ results given above to $n_c=4$.
\begin{remark} \label{extendto4} The conditions in the case of $n_c=4$ can be obtained by starting from a $3\times n_v$ matrix of the desired girth and using the connection between $C_{H_{4}}$ and $C_{H_{3}}.$

In Section~\ref{sec:mbyn}, we show how to use Lemma \ref{induction} efficiently to do this in the case $\girth(H)>8$, for which 
$C_H^2\triangle (C_H +I) =0$ must be satisfied, 
and, equivalently, 
\begin{align*}\left\{\begin{array}{l} (C_{H_3}^2+C_4C_4^\tr )\triangle (C_{H_3}+I) =0, \\ C_{H_3}C_4 \triangle C_4=0,\\C_4^\tr C_4 \triangle I=0.\end{array}\right.
\end{align*} This results in an  ``inductive" construction: we start from a $3\times n_v$ matrix of girth larger than 8 to insure that $C_{H_3}^2\triangle (C_{H_3}+I) =0$ and thus reduce  the above 3 conditions 
to only the ones that derive conditions on  exponents $k_l $. 
In~\cite{gomezfonseca2021necessary}, we provide all conditions for $n_c=4$ (derived using a more direct approach)  together with simulations of constructions using the algorithms.  
\end{remark} 
\subsection{Case of girth $\girth(H_{n_c})=2m> 12$} \label{sec:girth>12}
If we want girth larger than 12, we cannot  take $H_{n_c}$ to be composed solely of circulants, since it is well-known that a circulant lifting of a $2\times 3$ all-one protograph limits the girth to be 12, see, e.g., \cite{sv12}; therefore, we need to consider a matrix composed of permutation matrices  such that some are not circulant.  For $n_c = 3$, let $P_i,Q_i$ permutation matrices and 
\begin{equation}
H_3  = \begin{bmatrix} 1 & 1 & \cdots & 1\\ P_1 & P_2 & \cdots & P_{n_v}\\Q_1 & Q_2 & \cdots & Q_{n_v}\end{bmatrix},
\end{equation}
where we can take  $P_1=Q_1=1$ without loss of generality.

We start with the remark  that the parity-check matrix of every QC-LDPC code can be seen as a pre-lifted matrix with lifting factor $N=N_1N_2$ following a sequence of liftings, one of factor $N_1$ followed by a lifting of factor $N_2$. So one way to construct codes  of larger girth than 12 (or to increase its minimum distance) is to rewrite the parity-check  matrix of a QC-LDPC code of girth 12  as an equivalent pre-lifted matrix, and then  modify   some of the exponents to increase the girth and/or  the minimum distance. 
\begin{theorem} \label{thm:pre-lift} Every  QC-LDPC code with parity-check matrix $H$ of length $N=N_1N_2$ is equivalent to a pre-lifted QC-LDPC code of with pre-lift size (first lifting factor) $N_1$ and circulant size (second lifting factor) $N_2$.
\end{theorem} 
\begin{IEEEproof} 
We can transform each polynomial entry $g(x)= g_0(x^{N_1})+xg_1(x^{N_1})+\cdots + x^{N_1-1}g_{N_1-1} (x^{N_1})$ of $H$ into an $N_1\times N_1$ equivalent matrix, 
$$\begin{bmatrix} g(x) \end{bmatrix}= \begin{bmatrix} g_0(x)&xg_{N_1-1}(x)& xg_{N_1-2}(x) &\cdots & xg_{1}(x)\\ g_{1}(x)& g_0(x)&xg_{N_1-1}(x)&\cdots & xg_{N_1-2}(x)\\ \vdots&\vdots&\cdots &\vdots\\ g_{N_1-1}(x)& g_{N_1-2}(x) &g_{N_1-3}(x)&\cdots & g_0(x)\end{bmatrix}. $$
In the scalar matrix $H$ we can see this equivalence by performing a sequence of  column and row permutations (reordering of the columns and rows). We abuse the notation and use the equality sign between the two equivalent  representations (which result in  equivalent graphs). 
\end{IEEEproof} 
For example, if $N_1=2$ and $N=2N_2$, then the entries $x^{2a} =(x^2)^a$ and $x^{2a+1}=x (x^2)^a $ give the following transformations $$\begin{bmatrix} x^{2a} \end{bmatrix}= \begin{bmatrix} x^{a} &0\\0&x^a\end{bmatrix} \quad \text{ and } \quad \begin{bmatrix} x^{2a+1} \end{bmatrix}=\begin{bmatrix} 0& x^{a+1}\\x^a&0\end{bmatrix}. $$ 
\noindent  Similarly, if $N=3N_2$, the  equivalent code is $$\begin{bmatrix} x^{3a} \end{bmatrix}=\begin{bmatrix} x^{a} &0&0\\0&x^a&0\\0&0&x^a\end{bmatrix}, \quad \begin{bmatrix} x^{3a+1} \end{bmatrix}=\begin{bmatrix} 0& 0&x^{a+1}\\x^a&0&0\\0&x^a&0\end{bmatrix}, \quad \text{and}\quad \begin{bmatrix} x^{3a+2} \end{bmatrix}= \begin{bmatrix} 0& x^{a+1}&0\\0&0&x^{a+1}\\x^a&0&0\end{bmatrix},$$ with component matrices of size $N_2\times N_2$. 
In Appendix~\ref{pre-lift-observed-1}, we revisit Example~\ref{2by3} to show how Theorem \ref{thm:pre-lift} can be used to obtain  matrices of girth 24.   

\begin{example}\label{girth12-from 10-2}
In Appendix~\ref{pre-lift-observed-2}, we consider the  $(3,5)$-regular matrix $H_{3,g>8}$ from Example~\ref{girth12-from 10} of girth 10.  
 To improve performance, we first rewrite  the code  to display a  pre-lifted protograph with $N_1=2$ and then modify the exponents to achieve girth $12$ for the same code length as the single lifting of  $H_{3,g>10}$ from Example~\ref{girth12-from 10} (giving a code of length $1640$ in both cases). The simulated decoding performance of both the original codes, the pre-lifted code, and a random QC-LDPC code with similar parameters are provided in Section~\ref{sec:simulations}.
\end{example}
 
  As mentioned above,  the pre-lifted protograph must be free of a $2\times 3$ all-one matrix to achieve QC-LDPC matrices with girth larger than 12. In the next examples, we demonstrate how the (equivalent) pre-lift of earlier designs limit the girth to be 12 and how a pre-lift can be selected to avoid such limiting sub-structures. 
 \begin{example} \label{35unsuc} Consider the $(3,5)$-regular matrix $H_{3, g>10}$ of girth 12 constructed in Example~\ref{girth12-from 10}. Suppose that we write it as a $N_1=3$ prelift, described compactly as $$\begin{bmatrix} P_2&P_3&P_4&P_5\\Q_2&Q_3&Q_4&Q_5\end{bmatrix}=\left[\begin{array}{ccc|ccc|ccc|ccc} 
0&0&\matr{x}&    0&0&\boxed{\matr{x^3}}&   \matr{x^4}&0&0&   0&\boxed{x^7}&0 \\ 
1& 0&0&            x^2&0&0           &     0&x^4&0&          0&0&\matr{x^7}\\ 
0&1& 0&             0&x^2&0          &      0&0&x^4                &x^6&0&0\\ \hline
x^{22} &0&0             &0&0&\boxed{x^{36}}  &    0&0     &    x^{82}              &0&\boxed{x^{65}}&0         \\ 
0& x^{22}&0 &    \matr{x^{35}}&0&0 &     \matr{x^{81}}&0& 0         &0&0&x^{65}     \\
0&0&x^{22}&       0&\matr{x^{35}}&0&       0& \matr{x^{81}} &0      &x^{64}&0&0
\end{array}\right].$$
As can be seen, several 4-cycles exist in the submatrix above (one such example is highlighted by boxed values) that correspond to a $2\times 3$ submatrices with the identity matrices to the left (not shown). Hence, simply modifying exponents could not possibly increase the girth beyond 12 for any $N_2$.

One can also try to modify  the protograph  to avoid as many  $2\times 3$ all-one matrices as possible. The matrix below was modified from the above, using some non-circulant matrices for the pre-lift, but it still contains such $2\times 3$ submatrices (where entries involved are denoted with variables $A$, $B$, and $C$):
  $$ \begin{bmatrix} P_2&P_3&P_4&P_5\\Q_2&Q_3&Q_4&Q_5\end{bmatrix}= \left[\begin{array}{ccc|ccc|ccc|ccc} 
0&0&\matr{x}&    0&0&\matr{x^3}&  0 &\matr{x^{39}}&0&   0&x^{29}&0 \\ 
1& 0&0&            x^9&0&0           &     0&0&x^4&          0&0&\matr{x^{59}}\\ 
0&1& 0&             0&x^{17}&0          &      x^{11}  &0&0              &x^{71}&0&0\\ \hline
0&  x^{118}& 0          &0&0&x^{136} &    0&0     &    x^{290}              &x^{353}&0& 0        \\ 
0&0&x^{32} &   0 &\matr{x^{479}}&0 &     A&0& 0         &0&B& 0   \\
 x^{209}&0 &0&     C&0&0&       0& \matr{x^{800}} &0      &0&0&x^{-319}
\end{array}\right].$$
Again, any choice of circulants in those entries would limit the girth to 12 or less; however, setting $A= B=C=0$  (masking) eliminates the limiting structures and thus it is possible to further increase the girth. Indeed, masking and modifying the exponents as shown above gives a girth 14 irregular code for $N=891$.   Setting $A=x^{1199}, B=x^{1239}$, and $C=x^{-579}$ gives a $(3,5)$-regular matrix with girth 12, but where many  12-cycles  were eliminated by choosing the exponents to give an (irregular) code of girth 14. 
Both the irregular code of girth 14 and the regular code of girth 12  are simulated for $N=891$ (or length $n=13,365$) in Section \ref{sec:simulations}.
 \end{example}  
  
In Example \ref{35unsuc} we were not successful in obtaining a $(3,5)$-regular QC-LDPC matrix of girth 14, since it is not trivial to avoid the limiting sub-structures in the pre-lifted protograph without thought. We now show that the pre-lifted protograph can be chosen/designed in a deterministic way to avoid such structures. We note that starting from a known `good' code, e.g., one with a parity-check matrix of girth 12, provides a good base that can be modified relatively easily to increase the girth. Indeed, we see below that some good codes can be observed as derived from a  good pre-lifted protograph.

\begin{example} \label{ex:35g14}
We construct a matrix $H_3$ with submatrix $$\begin{bmatrix} P_2&P_3&P_4&P_5\\Q_2&Q_3&Q_4&Q_5\end{bmatrix}=\begin{bmatrix} x&x^7&x^{18}&x^{44}\\  x^{3} & x^{158}& x^{136}& x^{106}\end{bmatrix}. $$
This matrix has girth 12 for $N=279=3^2\cdot 31$.  The  expansion of $H_3$ as an equivalent code  with an observed 3-prelift  is  $$ \left[\begin{array}{ccc|ccc|ccc|ccc}
0&0&\matr{x}&    0&0&\matr{x^3} &\matr{x^{6}}&0&0&   0&x^{15}&0 \\ 
1& 0&0&            x^2&0&0           &     0&x^{6}& 0&         0&0&\matr{x^{15}}\\ 
0&1& 0&             0&x^{2}&0          &      0&0&x^{6}                &x^{14}&0&0\\ \hline
x&0& 0          &0&x^{53}& 0&    0&0     &    x^{46}              &0&0& x^{36}       \\ 
0&x & 0&  0 &0&\matr{x^{53}} &     \matr{x^{45}}&0& 0         &x^{35}&0& 0   \\
0&0 &x&       \matr{x^{52}}&0&0&       0& \matr{x^{45}} &0      &0&x^{35}&0
\end{array}\right],$$
where we note that the original matrix was carefully selected such that the pre-lift is free of $2\times 3$ all-one submatrices. The pre-lifted protograph of the above matrix 
corresponds  to 
$$\begin{bmatrix}  x&x&1&x^2\\ 1&x^2&x&x\end{bmatrix}$$ 
which  does not result in any $2\times 3$ all-one submatrix  (even though 4-cycles exist in the protograph) because each row multiplied by $x$ and $x^2$ does not overlap with the other rows in more than 2 positions. 

The following submatrix is obtained by modifying certain exponents in the above that participate in 12-cycles
 $$  \left[\begin{array}{ccc|ccc|ccc|ccc} 
0&0&\matr{x}&    0&0&\matr{x^3} &\matr{x^{6}}&0&0&   0&x^{15}&0 \\ 
1& 0&0&            x^5&0&0           &     0&x^{23}& 0&         0&0&\matr{x^{19}}\\ 
0&x^7& 0&             0&x^{11}&0          &      0&0&x^{29}                &x^{42}&0&0\\ \hline
x^{25}&0& 0          &0&x^{61}& 0&    0&0     &    x^{94}              &0&0& x^{153}       \\ 
0&x^{64} & 0&  0 &0&\matr{x^{180}} &     \matr{x^{239}}&0& 0         &x^{358}&0& 0   \\
0&0 &x^9&       \matr{x^{143}}&0&0&       0& \matr{x^{256}} &0      &0&x^{474}&0
\end{array}\right].$$
It has girth 14 for $N=752$  (code length $n=11,280$) and also, e.g.,  for $N=903$ (code length $n=13,545$) that we simulate in Section \ref{sec:simulations} to compare with codes of similar lengths from Example \ref{35unsuc}. 
\end{example}

 The final example demonstrates a construction of a girth 14 regular code obtained from a pre-lifted protograph of size $N_1=5$ that corresponds to a protograph of girth 6.   This choice ensures \emph{a priori} that the 5-cover does not have any  $2\times 3$ all-one submatrix.  

\begin{example} \label{structures}
The matrix $H_3$  with $$\begin{bmatrix} P_2&P_3&P_4&P_5\\Q_2&Q_3&Q_4&Q_5\end{bmatrix}=\begin{bmatrix} x&x^7&x^{18}&x^{44}\\ x^{32} & x^{54}& x^{141}& x^{133}\end{bmatrix}=$$ 
 {\tiny $$ \left[\begin{array}{ccccc|ccccc|ccccc|ccccc} 
0&0&0&  0&x&              0 &0&0&x^2&0&           0&0&x^4&0&0&   0&x^9&0&0&0   \\ 
  1& 0&0&0&0&              0 &0&0& 0&x^2&         0&0&0& x^4&0&   0&0&x^9&0&0  \\ 
  0&1& 0&0&0&              x & 0&0&0&0   &          0&0&0& 0& x^4 & 0&0&0&x^9&0   \\
  0&0&1&0&0 &              0&x&0 &0& 0&             x^3&0& 0&0&0&   0& 0&0&0&x^9  \\
0&0&0&1& 0&              0&0& x&0& 0&             0&x^3 &0& 0&0&    x^8&0&0&0&0 \\\hline
 0 &0&0&x^7&0&       0&x^{11}&0&0&0       &  0&0&0&  0&x^{29}&              0&0&x^{27}&0&0     \\ 
 0 &0&0& 0&x^7&      0&0&x^{11}&0&0      &  x^{28}& 0&0&0&0&           0&0&0& x^{27}&0 \\
 x^6 & 0&0&0&0&      0&0&0&x^{11}&0    &    0&x^{28}& 0&0&0&         0&0&0& 0& x^{27}   \\
 0&x^6&0 &0& 0&        0& 0&0&0&x^{11}  &   0&0&x^{28}&0&0 &          x^{26}&0& 0&0&0 \\
0&0& x^6&0& 0&       x^{10}&0&0&0&0      &     0&0&0&x^{28}& 0&            0&x^{26} &0& 0&0
\end{array}\right]$$}
\noindent has girth 12 for $N=245=5\cdot  49$, i.e., $N_1=5, N_2=49$.  
Note that the pre-lifted protograph of $H_3$ corresponds to
$$\begin{bmatrix} 1&1&1&1&1\\1& x&x^2&x^3&x^4\\1&x^2 &x^4&x&x^3\end{bmatrix}= \begin{bmatrix} 1&1&1&1&1\\1& x&x^2&x^3&x^4\\1&x^2 &x^4&x^6&x^8\end{bmatrix},$$ which has girth 6. So not only does it not have any $2\times 3$ all-one submatrix, it also does not have any $2\times 2$ all-one submatrix. 
Such a pre-lifted protograph will definitely give a girth 14 code and can also give a girth 16 because the matrix cannot include any submatrix $X$ below which is the structure shown in \cite{4276926} to restrict the girth to 14. In addition,  since the pre-lift has girth 6, no substructure $Y$ below  is present either, thus allowing a possible girth of 14, 16, and even 18, for which the forbidden structures are  $Z$ and $T$: 
$$X=\begin{bmatrix} 1&1&1\\1&1&0\\1&0&1\end{bmatrix}, Y=\begin{bmatrix} 1&1\\1&1\end{bmatrix}, Z=\begin{bmatrix} 1&1&1&1\\1&1&0&0\\0&0&1&1\end{bmatrix}, T=\begin{bmatrix} 1&1&1&0\\1&1&0&1\\0&0&1&1\end{bmatrix}.$$ 
We modified the above exponents to obtain a girth 14 matrix for $N=605$ given by
$$\begin{bmatrix} P_2&P_3&P_4&P_5\\Q_2&Q_3&Q_4&Q_5\end{bmatrix}=$$
{\tiny $$\left[\begin{array}{ccccc|ccccc|ccccc|ccccc} 
   0&0&0&  0&x&              0 &0&0&x^2&0&           0&0&x^4&0&0&   0&x^9&0&0&0   \\ 
  1& 0&0&0&0&              0 &0&0& 0&x^2&         0&0&0& x^4&0&   0&0&x^9&0&0  \\ 
 0&x^3& 0&0&0&              1 & 0&0&0&0   &          0&0&0& 0& x^8 & 0&0&0&x&0   \\
   0&0&x^9&0&0 &              0&x^{13}&0 &0& 0&             x^{19}&0& 0&0&0&   0& 0&0&0&x^{23}  \\
0&0&0&x^7& 0&              0&0& x^{19}&0& 0&             0&x^{34} &0& 0&0&    x^{44}&0&0&0&0 \\\hline
  0 &0&0&x^{29}&0&       0&x^{40}&0&0&0       &  0&0&0&  0&x^{79}&              0&0&x^{99}&0&0     \\ 
0 &0&0& 0&x^{29}&      0&0&x^{54}&0&0      &  x^{115}& 0&0&0&0&           0&0&0& x^{135}&0 \\
 x^{23} & 0&0&0&0&      0&0&0&x^{73}&0    &    0&x^{129}& 0&0&0&         0&0&0& 0& x^{215}   \\
 0&x^{55}&0 &0& 0&        0& 0&0&0&x^{145}  &   0&0&x^{209}&0&0 &          x^{313}&0& 0&0&0 \\
0&0& x^{301}&0& 0&       x^{356}&0&0&0&0      &     0&0&0&x^{432}& 0&            0&x^{512} &0& 0&0
\end{array}\right]$$}
Such modification can be achieved relatively easily by masking and then unmasking circulants one by one and choosing them such that the girth is 14 (or larger as desired).
 \end{example} 
  
 \begin{theorem}\label{prelift-girth6}
Let $B$ be an $(n_c,n_v)$-regular $n_cN_1\times n_vN_1$  parity-check matrix  of a protograph-based QC-LDPC code of girth 6. Then there exist a lifting factor $N$ for which $B$  can be lifted to obtain a QC-LDPC code with parity-check matrix $H_{n_c}$ of  girth 14,  16, or  18, as desired.   
\end{theorem} 
\begin{IEEEproof} From \cite{4276926},  we observe that all the structures that need to be avoided in order to allow for girth 14, 16, and 18 contain a $2\times 2$ all-one submatrix (they have girth 4).  These structures are discussed in Example~\ref{structures}. Since $B$ is of girth 6 and it acts as a pre-lifted protograph used on an $n_c\times n_v$ all-one protograph to obtain $H_{n_c}$, then $H_{n_c}$ cannot  possibly contain any of these substructures. 
\end{IEEEproof} 
 
 \begin{remark}  
 We remind the reader that for girth 14 and above we use the computer to search for the next good value and used a value of $N$ large enough to allow such a value.   Algorithms like the ones we presented in Section \ref{sec:3byn-girth} for girth up to 12 could be developed,  but due to the fact that each protograph needs to be considered separately, it will answer only this case rather than allow for a general algorithm like those earlier. This could nevertheless be attractive if the protograph has been optimized; we show how this could be done for the  NASA CCSDS protograph in Section \ref{sec:multi}. \end{remark}

\begin{theorem}\label{prelift-girth8}
Let $B$ be an $(n_c,n_v)$-regular $n_cN_1\times n_vN_1$ parity-check matrix of a protograph-based QC-LDPC code of girth 8. Then there exist lifting factors $N$ for which $B$  can be lifted to obtain a QC-LDPC code with parity-check matrix $H_{n_c}$ of  girth 20 or 22, as desired.   
\end{theorem} 
\begin{IEEEproof} From paper~\cite{4276926} and Example~\ref{structures}, most of the forbidden structures for girth 20 and 22 contain a $2\times 2$ all-one matrix. There is one structure for girth 20 and one for girth 22 that has girth 6.  Taking $B$ of girth 8 guarantees that these structures are not present in the lifted $H_{n_c}$ from $B$. 
\end{IEEEproof} 
\begin{remark}
 Theorems~\ref{prelift-girth6} and \ref{prelift-girth8}  are only sufficient but not necessary. Example \ref{ex:35g14} does not satisfy these theorems but demonstrates that a $3\times 5$ matrix of girth 14 can be obtained from the all-one matrix pre-lifted with $N_1=3$  provided that the pre-lifted protograph excludes the limiting structure.
\end{remark}
\section{Protographs with multiple edges}\label{sec:multi}
In this section, we briefly address the case in which the original protograph is not all-one, to demonstrate how the theory and techniques can be extended to such protographs. In particular, we revisit the earlier example of the NASA CCSDS protograph that was discussed in the context of sums of permutation matrices in Examples \ref{NASA-example} and \ref{NASA-example-cont}, and now connect to the  perspectives developed in Section \ref{sec:3bynv}.

We proceed then to consider a $4\times 8$ protograph that contains entries equal to 2 (multiple edges) and zero matrices as shown in \eqref{CCSDS}. Without loss of generality, we can let $P_8=Q_5=R_6=S_7=I$ and assume the following  reduced matrix because, by multiplying rows/columns  with permutation matrices, we obtain equivalent graphs and equivalent corresponding codes: 
$$H\defeq  \left[\begin{array}{c|c}  A& B\end{array}\right] \defeq   \left[\begin{array}{c|c} \underbrace{\begin{matrix} I+P_1 &P_2&P_3&P_4\\Q_1&I+Q_2 & Q_3&Q_4\\R_1&R_2&I+R_3& R_4\\S_1&S_2&S_3& I+S_4\end{matrix}}_{A}&\underbrace{\begin{matrix} 
 0&P_6 &P_7&P_8\\ Q_5&0& Q_7&Q_8\\ R_5&R_6&0& R_8\\S_5&S_6& S_7&0\end{matrix}}_{B}
\end{array}\right] =$$ $$\left[\begin{array}{c|c}\begin{matrix}1+x^{a_1} &x^{a_2}&x^{a_3}&x^{a_4}\\ x^{b_1}&1+ x^{b_2}& x^{b_3} & x^{b_4}\\
x^{c_1}& x^{c_2}&1+ x^{c_3}&  x^{c_4}\\ x^{d_1}& x^{d_2}& x^{d_3}& 1+x^{d_4}\end{matrix} &\begin{matrix} 
 0&x^{a_6} &x^{a_7}&x^{a_8}\\  x^{b_5}&0&  x^{b_7} & x^{b_8}\\   x^{c_5}& x^{c_6}&0&  x^{c_8}\\ x^{d_5}&x^{d_6}&  x^{d_7}&0
\end{matrix}\end{array}\right]  $$

In this case, $HH^\tr =AA^\tr+BB^\tr=  8I+ \underbrace{C_{A}+C_{B}}_{C_H}$ 
 where, unlike in the case of protographs without multiple edges, $C_A=(C_{A, ij})_{i,j\in[4]}$ and, therefore  $C_H= C_{A}+C_{B}$,   has non-zero entries on the main diagonal, i.e., $C_{A, ii}\neq 0$,  and $C_{H, ii}\neq 0$, respectively. The component matrices $C_{A,ij}$ for  $C_{A}$ are computed as
   \begin{align*} 
 \begin{matrix}\matr{C_{A, 11}} \defeq x^{a_1}+x^{-a_1}, \matr{C_{A, 22}} \defeq x^{b_2}+x^{-b_2}, \matr{C_{A, 33}} \defeq x^{c_3}+x^{-c_3}, \matr{C_{A, 44}} \defeq x^{d_4}+x^{-d_4},\\
 \matr{C_{A, 12}}\defeq C_{A, 21}^\tr 
 \defeq x^{-b_1}+x^{a_2}+ \sum\limits_{j=1}^{4}  x^{a_j-b_j},  \matr{C_{A, 13}}\defeq C_{A, 31}^\tr 
 \defeq  x^{-c_1}+x^{a_3}+ \sum\limits_{j=1}^{4}  x^{a_j-c_j}, \\
 \matr{C_{A, 14}}\defeq C_{A, 41}^\tr 
 \defeq   x^{-d_1}+x^{a_4}+ \sum\limits_{j=1}^{4}  x^{a_j-d_j}, 
 \matr{C_{A, 23}}\defeq C_{A, 32}^\tr 
  \defeq  x^{-c_2}+x^{b_3}+ \sum\limits_{j=1}^{4}  x^{b_j-c_j}, \\
 \matr{C_{A, 24}}\defeq C_{A, 42}^\tr 
 \defeq  x^{-d_2}+x^{b_4}+ \sum\limits_{j=1}^{4}  x^{b_j-d_j}, 
 \matr{C_{A, 34}}\defeq C_{A, 43}^\tr 
 \defeq  x^{-d_3}+x^{c_4}+ \sum\limits_{j=1}^{4}  x^{c_j-d_j} , \end{matrix}
  \end{align*} 
and the component matrices $C_{B,ij}$ for $C_B$ are 
\begin{align*} 
 \begin{matrix}\matr{C_{B, 11}}\defeq 0,  \matr{C_{B, 22}} \defeq 0, \matr{C_{B, 33}}\defeq 0,  \matr{C_{B, 44}} \defeq 0\\
 \matr{C_{B, 12}}\defeq C_{B, 21}^\tr \defeq  \sum\limits_{j=7,8}  x^{a_j-b_j},  \matr{C_{B, 13}}\defeq C_{B, 31}^\tr \defeq  \sum\limits_{j=6,8}  x^{a_j-c_j}, \matr{C_{B, 14}}\defeq C_{B, 41}^\tr \defeq    \sum\limits_{j=6,7} x^{a_j-d_j},  \\
 \matr{C_{B, 23}}\defeq C_{B, 32}^\tr  \defeq  \sum\limits_{j=5,8} x^{b_j-c_j},
 \matr{C_{B, 24}}\defeq C_{B, 42}^\tr \defeq  \sum\limits_{j=5,7}  x^{b_j-d_j}, 
 \matr{C_{B, 34}}\defeq C_{B, 43}^\tr \defeq    \sum\limits_{j=5,6}  x^{c_j-d_j}. \end{matrix}
  \end{align*}  
Therefore, {\small $$C_H= 
\begin{bmatrix} x^{-a_1}+x^{a_1} &  x^{-b_1}+x^{a_2}+ \sum\limits_{{j\in[8]}\atop {j\neq 5,6}}  x^{a_j-b_j}   & 
x^{-c_1}+x^{a_3}+ \sum\limits_{{j\in[8]}\atop {j\neq 5,7}}  x^{a_j-c_j} &  x^{-d_1}+x^{a_4}+ \sum\limits_{{j\in[8]}\atop {j\neq 5,8}}  x^{a_j-d_j} \\
x^{-a_2}+x^{b_1}+ \sum\limits_{{j\in[8]}\atop {j\neq 5,6}}  x^{-a_j+b_j} & x^{-b_2}+x^{b_2} &x^{-c_2}+x^{b_3}+ \sum\limits_{{j\in[8]}\atop {j\neq 6,7}}  x^{b_j-c_j} &  x^{-d_2}+x^{b_4}+ \sum\limits_{{j\in[8]}\atop {j\neq 6,8}}  x^{b_j-d_j} & \\
x^{-a_3}+x^{c_1}+ \sum\limits_{{j\in[8]}\atop {j\neq 5,7}}  x^{-a_j+c_j}&x^{-b_3}+  x^{c_2}+ \sum\limits_{{j\in[8]}\atop {j\neq 6,7}}  x^{-b_j+c_j} &  x^{-c_3}+x^{c_3} &x^{-d_3}+x^{c_4}+ \sum\limits_{{j\in[8]}\atop {j\neq 7,8}}  x^{c_j-d_j} \\  
x^{-a_4}+x^{d_1}+ \sum\limits_{{j\in[8]}\atop {j\neq 5,8}}  x^{-a_j+d_j} &x^{-b_4}+ x^{d_2}+ \sum\limits_{{j\in[8]}\atop {j\neq 6,8}}  x^{-b_j+d_j} &x^{-c_4}+x^{d_3}+ \sum\limits_{{j\in[8]}\atop {j\neq 7,8}}  x^{-c_j+d_j} &x^{-d_4}+x^{d_4}
\end{bmatrix} $$}

Note that Theorem~\ref{adjacent-cond-new}  still holds, so in computing $B_t$ we only need to consider  powers of $C_H$.
For example, if we desire to avoid 4-cycles then $C_H\triangle I=0$ must hold, which is equivalent to 
$$ \left\{\begin{matrix}  C_{A,ii} \triangle I =0, \text{ for all }  i\in[4], \\
C_{A,ij} +C_{B,ij}\triangle 0 =0,  \text{ for all }  i,j\in[4], i\neq j.
\end{matrix}\right. $$ 
Equivalently, no value in the set $\{ 2a_{1}, 2b_{2}, 2c_{3},2d_{4}\} $ is 0 modulo $N$,  and each of the sets below are of maximal size
$$\{-b_1, a_2,  a_j-b_j, j \in [8], j\neq 5,6\} , \{-c_1, {a_3}, {a_j-c_j},  j \in [8], j\neq 5,7\}, $$
$$\{-d_1,a_4, {a_j-d_j}, j \in [8], j\neq 5,8\}, \{-c_2,{b_3}, {b_j-c_j}, j \in [8], j\neq 6,7\}, $$ $$\{-d_2, {b_4}, {b_j-d_j}, j \in [8], j\neq 6,8\}, \{-d_3, {c_4}, {c_j-d_j}, j \in [8], j\neq 7,8\}. $$

\noindent A fast algorithm similar to those in Section \ref{sec:3byn-girth} can be created to construct matrices $H$ satisfying the conditions above. 

\begin{example} \label{multipleedges-example} We revisit the matrix $H_{(128,64)}$ from Example~\ref{NASA-example}, this time as a $4\times 8$ protograph-based matrix of the form above,  in order to compute the matrix $C_{H_{(128,64)}}$ and show that it satisfies the conditions for the matrix $H_{(128,64)}$ to have girth 6.  
Indeed, the matrix 
 $$H_{(128,64)}\defeq \left[\begin{array}{c|c}\begin{matrix}1+x^{7} &x^{2}&x^{14}&x^{6}\\ x^{6}&1+ x^{15}& 1 & x\\
x^{4}& x&1+ x^{15}&  x^{14}\\ 1& x& x^{9}& 1+x^{13}\end{matrix} &\begin{matrix} 
 0&1&x^{13}&1\\  1&0&  1& x^{7}\\   x^{11}& 1&0&  x^{3}\\ x^{14}&x&  1&0
\end{matrix}\end{array}\right]  $$
with $N=16$,  satisfies all the conditions above (these can be easily checked by hand), and hence it has girth 6. 
\end{example} 

A similar approach and theory to that developed above for the NASA CCSDS protograph can be developed for arbitrary protographs with multiple edges.

\section{Simulation results}\label{sec:simulations}
To verify the performance of the constructed codes, computer simulations were performed assuming binary phase shift keyed (BPSK) modulation and a binary-input additive white Gaussian noise (AWGN) channel. The sum-product message passing decoder was allowed a maximum of $100$ iterations and employed a syndrome-check based stopping rule. 

\begin{figure}[h]
\begin{center}
\includegraphics[width=4in]{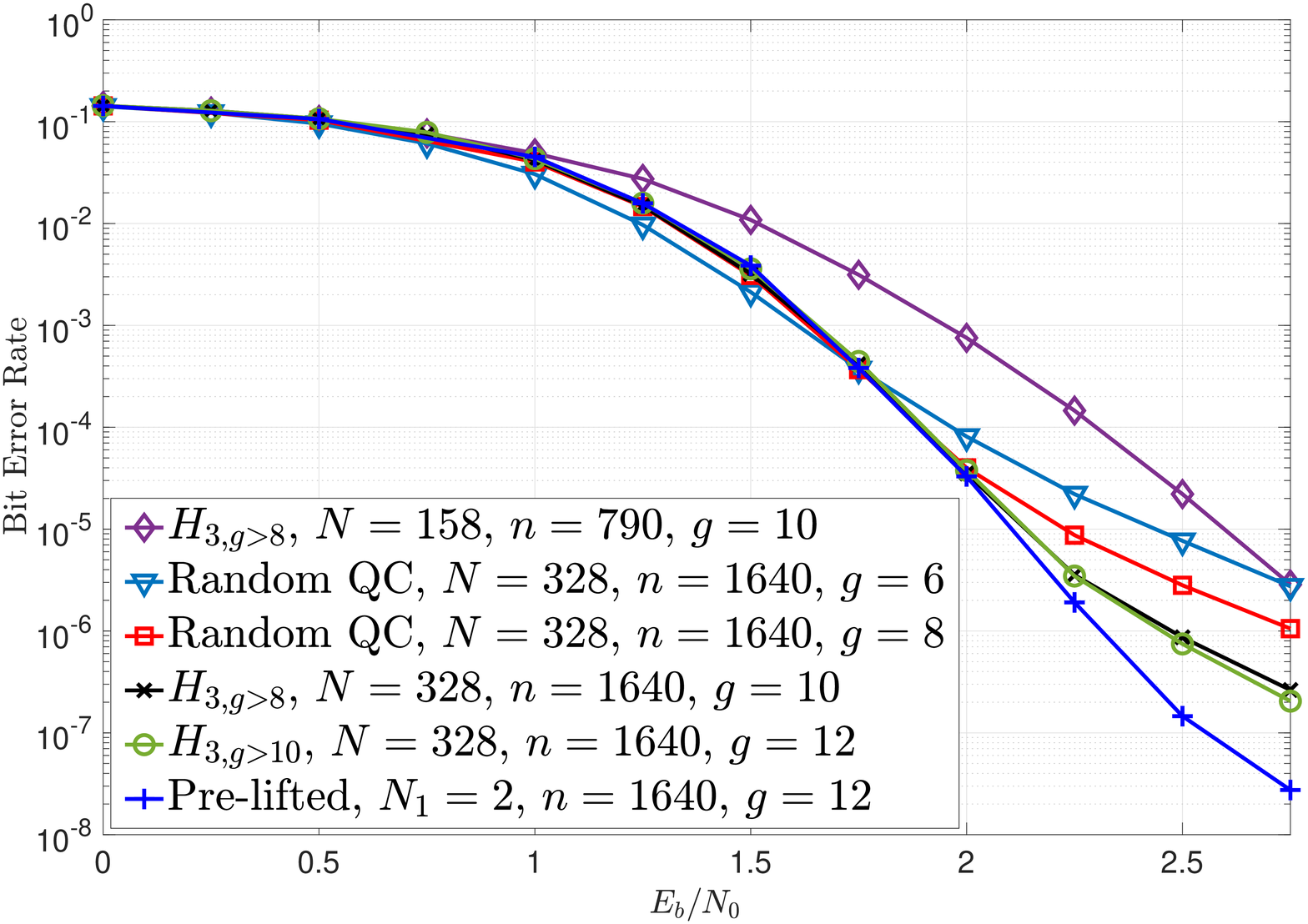}\vspace{-4mm}
\end{center}
\caption{Simulated decoding performance in terms of BER for the $R=2/5$ QC-LDPC codes from Examples \ref{girth12-from 10} and \ref{girth12-from 10-2}.}\label{fig:short}
\end{figure}

In Fig.~\ref{fig:short}, we plot the bit error rate (BER) for the $R\approx2/5$, $(3,5)$-regular QC-LDPC codes from Example \ref{girth12-from 10}. We show the performance of two codes of length $n=790$ ($N=158$) and $n=1640$ ($N=328$) derived from $H_{3,g>8}$, both with girth 10, the performance of the code of length $n=1640$ ($N=328$) derived from $H_{3,g>10}$ with girth 12, and two random QC liftings of the all-ones $3\times 5$ protograph with the same length ($N=328$) and respective girths of 6 and 8. We observe that the larger lifting factor results in an improved waterfall as expected and that the error floors of the large girth codes are lower than the random code.\footnote{An additional example comparing single lifts to random codes with various girth was presented in \cite{sm21}.} Finally, also shown is the pre-lifted version of the code from Example \ref{girth12-from 10-2} and Appendix~\ref{pre-lift-observed-2} with $N_1=2$ and $N_2=164$, which results in a $(3,5)$-regular code of length $n=1640$. In this construction, some exponents were modified to break some circular sub-structures and we see that the error floor is lowered compared to the single lifts.

\begin{figure}[h]
\begin{center}
\includegraphics[width=4in]{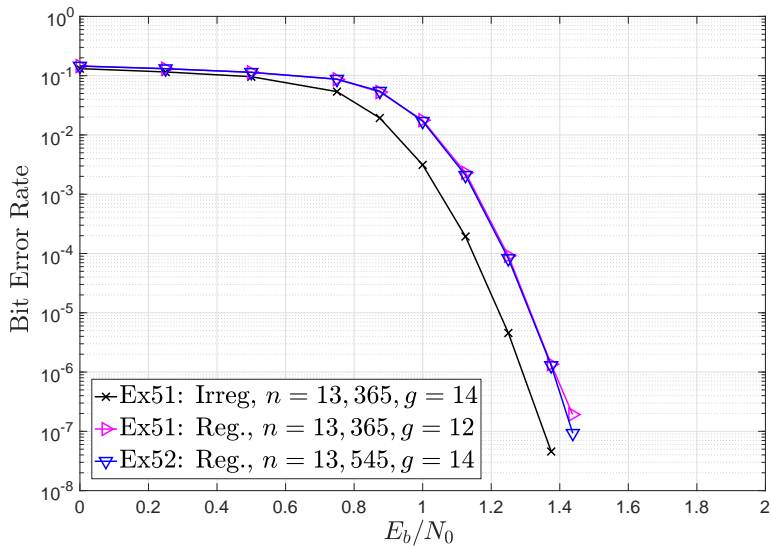}\vspace{-4mm}
\end{center}
\caption{Simulated decoding performance in terms of BER for the $R=2/5$ QC-LDPC codes from Examples \ref{35unsuc} and \ref{ex:35g14}.}\label{fig:comp}
\end{figure}

In Fig. \ref{fig:comp}, we plot the BER for the $R\approx2/5$ QC-LDPC codes with longer block lengths from Examples \ref{35unsuc} and \ref{ex:35g14}. We remark that these high girth codes display no indication of an error-floor, at least down to a BER of $10^{-7}$. The regular codes from Example \ref{35unsuc} (reduced multiplicity of 12 cycles) and \ref{ex:35g14} (with girth 14) have similar performance in the simulated range, but we anticipate deviation at higher SNRs where the 12-cycles are involved in trapping sets. For reference, the iterative decoding threshold for $(3,5)$-regular LDPC codes is 0.96dB \cite{ru08}. The irregular code of girth 14 is shown to outperform the regular codes in the simulated range. We remind the reader that the irregular code was obtained by masking some circulants from the pre-lifted code to increase the girth. Such a strategy can yield good optimized irregular LDPC codes. 

\section{Concluding Remarks}\label{sec: conclusion} In this paper we provided a unifying framework under which all constructions of girth between 6 and 12 can be included.  Towards this goal, we gave necessary and sufficient conditions for the Tanner graph of a protograph-based QC-LDPC code to have girth between 6 and 12. We also showed how these girth conditions can be used to write fast  algorithms to construct such codes and how to employ a double graph-lifting procedure,  called pre-lifting,  in order to exceed girth 12. 
We   showed that the cases of variable node degrees $n_c=2$, $3$, and $4$ that we consider in this paper are not just particular cases, but provide the girth framework for the $n_c\times n_v$ all-one protograph, for {\em all} $n_c\geq 2$. 
We also presented a new perspective on $n_cN\times n_vN$ LDPC protograph-based parity-check matrices  by viewing them as $n_cN$ rows of a parity-check matrix equal to the sum of certain $n_vN\times n_vN$ permutation matrices and highlighted an important connection between $n_c\times n_v$ protographs, for any $n_c\geq 2$,  and protographs with  $n_c=2$. Finally, we exemplifed how the results and methodology can be used and adapted to analyze the girth of the Tanner graph of any parity-check matrix 
on an irregular, multi-edge protograph  of the NASA CCSDS LDPC  code. 

\appendices\label{sec:appendices}
\section{Example~\ref{2by3} revisited} \label{pre-lift-observed-1}

In order to exemplify Theorem \ref{thm:pre-lift}, we revisit Example~\ref{2by3} and show how we obtained the matrices of girth 24.   Suppose that  a $2\times 3$ protograph based code has $I+P_2+P_3=1+x^2+x^3$ of  girth 6 (it satisfies the conditions  of girth 6) or, equivalently, $\girth(H)=12$.  We rewrite this as $$1+x^2+x^3 = \begin{bmatrix} 1+x & x^2\\x& 1+x  \end{bmatrix} =I+ \begin{bmatrix} x & 0\\0& x  \end{bmatrix}+ \begin{bmatrix}0 & x^2\\x& 0  \end{bmatrix}, \quad P_2= \begin{bmatrix} x & 0\\0& x  \end{bmatrix}, P_3= \begin{bmatrix}0 & x^2\\x& 0  \end{bmatrix}. $$ 

\noindent We now slightly modify one entry in this quasi-cyclic parity-check matrix, enough to break the equivalence to the cyclic code. For example, the $18\times 18$ matrix 
$$I+P^\prime_2+P_3\defeq \begin{bmatrix} 1+x & x^2\\x& 1+x^5  \end{bmatrix}, \quad P^\prime_2 = \begin{bmatrix} x &0\\0&x^5  \end{bmatrix}, P_3 = \begin{bmatrix} 0 &x^2\\x&0  \end{bmatrix}$$ has  an associated graph with girth 8.  Note that we had to increase the size of the circulant matrices in order to observe an increase in girth. It follows that the  LDPC code with $36\times 54$ parity-check matrix
$$
\begin{bmatrix} \matr{I} &\matr{I} &\matr{I}\\ \matr{I}& \matr{P^\prime_2}&\matr{P_3}\end{bmatrix}  =\left[\begin{array}{c|c|c} \matr{I} &\matr{I} &\matr{I}\\ \hline\matr{I}&\begin{matrix} x & 0\\0& x^5  \end{matrix}&\begin{matrix} 0 & x^2\\x& 0  \end{matrix} \end{array}\right] =\left[\begin{array}{c|c|c} \begin{matrix} 1 & 0\\0& 1  \end{matrix} &\begin{matrix} 1 & 0\\0& 1  \end{matrix} &\begin{matrix} 1 & 0\\0& 1  \end{matrix}\\ \hline\begin{matrix} 1 & 0\\0& 1  \end{matrix}&\begin{matrix} x & 0\\0& x^5  \end{matrix}&\begin{matrix} 0 & x^2\\x& 0  \end{matrix} \end{array}\right]  $$
has girth 16. 

Similarly, we can start from the cyclic code of length $21$ with the same parity-check matrix $1+x^2+x^3$.  We reorder the rows and the columns of the parity-check matrix or, equivalently, make the replacements such that
 $$I+P_2+P_3= \begin{bmatrix} 1+x& x &0\\ 0 &1+x&x \\ 1& 0& 1+x\end{bmatrix}, \quad  P_2=\begin{bmatrix}x&0&0 \\0&x&0\\ 0&0&x\end{bmatrix}, P_3=  \begin{bmatrix}0&x&0 \\0&0&x\\ 1&0&0\end{bmatrix}.$$ We  modify it as $$I+P^\prime_2+P^\prime_3\defeq  \begin{bmatrix} 1+x& x &0\\ 0 &1+x^{13} &x^2 \\ x& 0& 1+x^7\end{bmatrix}, \quad  P'_2= \begin{bmatrix}x&0&0 \\0&x^{13}&0\\ 0&0&x^7\end{bmatrix}, Q'_2=  \begin{bmatrix}0&x&0 \\0&0&x^2\\ x&0&0\end{bmatrix},$$  to obtain girth  8 for a circulant size $N=7$, girth 10  if the size of the circulant  is increased to $N=11$, and girth 12  if the size is increased to $N=31$. 
%
Therefore, the  corresponding parity-check matrix  $H$ has girth 24 for $N= 31$. 

\section{Example~\ref{girth12-from 10}  revisited}\label{pre-lift-observed-2}
 We consider the  $(3,5)$-regular matrix $H_{3,g>8}$ from Example~\ref{girth12-from 10} of girth 10.  
 To improve performance, we rewrite  the codes we constructed  to display a  pre-lifted protograph with $N_1=2$ 
$$H_{3,g>8}=
\left[\begin{array}{cc|cc|cc|cc|cc} \matr{1} & 0& \matr{1} &0& \matr{1}&0& \matr{1}&0 &1&0 \\ 
0& \matr{1} & 0& \matr{1} &0& \matr{1}&0& \matr{1}&0&1\\\hline
 \matr{1}& 0&0&\matr{x}&0&\matr{x^4}&x^6&0&x^{10}&0\\ 
 0& \matr{1}& 1& 0&\matr{x^3}&0&0&\matr{x^6}&0&x^{10}\\  \hline
 \matr{1}& 0&\matr{x^{33}}&0&\matr{x^{53}}&0&\matr{x^{122}}&0&x^{97}&0\\ 
 0& \matr{1}& 0&x^{33}& 0&\matr{x^{53}}&0&x^{122}&0&\matr{x^{97}}\end{array}\right].$$ 
 
 \noindent The following matrix was modified in two entries, such that the $3\times4$ submatrices do not have all permutation matrices circulant (and hence commutative) and thus they can  observe an increase in minimum distance and in girth. We need to modify at least one  of every group of 4.  The matrix
$$H_{3,g>10}= \left[\begin{array}{cc|cc|cc|cc|cc} \matr{1} & 0& \matr{1} &0& \matr{1}&0& \matr{1}&0 &1&0 \\ 
0& \matr{1} & 0& \matr{1} &0& \matr{1}&0& \matr{1}&0&1\\\hline
 \matr{1}& 0&0&\matr{x}&0&\matr{x^4}&x^6&0&x^{10}&0\\ 
 0& \matr{1}& 1& 0&\matr{x^3}&0&0&\matr{x^6}&0&x^{10}\\  \hline
 \matr{1}& 0&\matr{x^{33}}&0&\matr{x^{53}}&0&\matr{x^{122}}&0&x^{97}&0\\ 
 0& \matr{1}& 0&x^{33}& 0&\matr{x^{53}}&0&x^{93}&0&\matr{x^{122}}\end{array}\right] $$ 
 has girth 10 for $N=123$   and girth 12 for $N=164$. Simulation results for the second code are provided in Section \ref{sec:simulations}.

\bibliographystyle{IEEEtran}

\end{document}